%% file: neunern_main.tex
\documentclass[review,authoryear,10pt]{elsarticle}

\usepackage{lineno,hyperref}
\modulolinenumbers[0]
\usepackage{fullpage}
\usepackage{amsmath}
\usepackage{amssymb}
\usepackage{longtable}
\usepackage{booktabs}
\usepackage{tabularx}
\usepackage{multirow}
\usepackage{diagbox}
\usepackage{soul}
\usepackage{amsmath}
\usepackage{enumitem}  
\usepackage[labelfont=bf]{caption}
\usepackage{placeins}
\usepackage{subcaption}
\usepackage{verbatim}
\usepackage{rotating}
\usepackage[table,xcdraw]{xcolor}
\usepackage[colorinlistoftodos,prependcaption]{todonotes}
\usepackage{regexpatch}
\usepackage[T1]{fontenc}
\makeatletter
\xpatchcmd{\@todo}{\setkeys{todonotes}{#1}}{\setkeys{todonotes}{inline,#1}}{}{}
\makeatother
\journal{Elsevier}
\usepackage[utf8]{inputenc}
\usepackage{siunitx}

\newlength{\twosubht}
\newsavebox{\twosubbox}
\newcommand{\NineEuroTicket}{\mbox{9-Euro-Ticket}}

\usepackage[acronym]{glossaries} 
\newacronym{pt}{PT}{public transport}
\newacronym{mid}{MiD}{Mobilität in Deutschland}

\begin{document}

\begin{frontmatter}

\title{Germany's nationwide travel experiment in 2022: public transport for 9 Euro per month - First findings of an empirical study}
%\tnotetext[mytitlenote]{Fully documented templates ae available in the elsarticle package on \href{http://www.ctan.org/tex-archive/macros/latex/contrib/elsarticle}{CTAN}.}

\author[tumvt]{Allister Loder\corref{mycorrespondingauthor}}
\author[eco]{Fabienne Cantner}
\author[ftm]{Lennart Adenaw}
\author[ftm]{Nico Nachtigall}
\author[ftm]{David Ziegler}
\author[ftm]{Felix Gotzler}
\author[ttt]{Markus B. Siewert}
\author[pol]{Stefan Wurster}
\author[eco]{Sebastian Goerg}
\author[ftm]{Markus Lienkamp}
\author[tumvt]{Klaus Bogenberger}

\cortext[mycorrespondingauthor]{Corresponding author \\ E-mail address: allister.loder@tum.de }

\address[tumvt]{Chair of Traffic Engineering and Control, 
TUM School of Engineering and Design, Technical University of Munich }
\address[ftm]{Chair of Automotive Technology, 
TUM School of Engineering and Design, Technical University of Munich }

\address[eco]{Professorship of Economics, 
TUMCS for Biotechnology and Sustainability \& TUM School of Management, Technical University of Munich }

\address[ttt]{TUM Think Tank, 
Munich School of Politics and Public Policy}

\address[pol]{Professorship of Policy Analysis, 
Munich School of Politics and Public Policy, TUM School of Social Sciences and Technology, Technical University of Munich}

\begin{abstract}
In spring 2022, the German federal government agreed on a set of policy measures that aimed at reducing households' financial burden resulting from a recent price increase, especially in energy and mobility. These included among others, a nationwide public transport ticket for 9~Euro per month for three months in June, July, and August 2022. In transport policy research this is an almost unprecedented behavioral experiment. It allows us to study not only behavioral responses in mode choice and induced demand but also to assess the effectiveness of these instruments. We observe this natural experiment with a three-wave survey and a smartphone-based travel diary with passive tracking on an initial sample of 2,261 participants with a focus on the Munich metropolitan region. This area is chosen as it offers a variety of mode options with a dense and far-reaching public transport network that even provides good access to many leisure destinations. The app has been providing data from 756 participants until the end of September, the three-wave survey by 1,402, and the app and the three waves by 637 participants. In this paper, we report on the study design, the recruitment and study participation as well as the impacts of the policy measures on the self-reported and app-observed travel behavior; we present results on consumer choices for a successor ticket to the \NineEuroTicket{} that started in May 2023. We find a substantial shift in the modal share towards public transport from the car in our sample during the \NineEuroTicket{} period in travel distance (around 5~\%) and in trip frequency (around 7~\%). The mobility outcomes of the \NineEuroTicket{} however provide evidence that cheap public transport as a policy instrument does not suffice to incentive sustainable travel behavior choices and that other policy instruments are required in addition. 

\end{abstract}


\end{frontmatter}

\clearpage

\section{Introduction}

In transport policy research, it is quite unlikely to perform real-world population-scale experiments on travel behavior. There are few notable exceptions: strikes suddenly make one  alternative mode unavailable \citep[e.g.,][]{anderson_subways_2014,adler_does_2016,bauernschuster_when_2017,larcom_benefits_2017}, a global pandemic changes travelers' preferences for traveling \citep[e.g.,][]{molloy_observed_2021,eisenmann_transport_2021,christidis_challenges_2022}, or a bridge collapse altering the provided transportation network \citep[e.g.,][]{zhu_traffic_2010}. In 2022, the German federal government created a population-scale natural experiment by providing a drastically discounted nationwide travel pass in response to the cost-of-living crisis caused by the geopolitical crisis in Ukraine that started in February 2022 \citep{bundesfinanzministerium_masnahmenpaket_2022}. The discounted nationwide flat-fare travel pass, called the \NineEuroTicket{}, priced at 9~Euro (approx. 9~USD in 2022), provided its holders to travel almost fare-free during June, July, and August 2022 all across the nation using all regional, local, and urban public transportation; the \NineEuroTicket{} was not valid for long-distance passenger services (e.g., high-speed rail services like ICE, TGV). Travelers could take their children up to six years with them free of charge. The \NineEuroTicket{} was also a measure to promote using public transport after the COVID-19 pandemic, where its usage declined drastically in Germany \citep{kolarova_analysing_2021} and did not fully recover by early 2022. 

The \NineEuroTicket{} rapidly became a ``darling'' of the public, and it attracted much interest. Various studies had been set up,  collecting their own survey data \citep[e.g.,][]{kramer_9-euro-ticket_2022,deutsche_bahn_ag_abschlussbericht_2022} or using third-party mobile phone data \citep[e.g.,][]{gaus_9-euro-ticket_2023} to interfere impacts on travel behavior and customer preferences. It had to be noted that the government announced various measures to ease the cost-of-living crisis. Here, another transport-related measure was a fuel excise tax cut of 29.55~Euro-Cent per liter of gasoline and for 14.04~Euro-Cent per liter of diesel. However, due to market price fluctuations, this price cut was not felt much by drivers. Consequently, the only transport-related intervention of the government that played a decisive role in decision-making was the \NineEuroTicket{}.

This natural experiment is promising from a travel behavior and transport policy perspective as new policy instruments are required to achieve the needed decarbonization of the transportation sector, which includes making sustainable travel choices more attractive \citep{lindsey_addressing_2020,axsen_crafting_2020}. At the behavioral level, the \NineEuroTicket{}  allows to reveal mode choice preferences under such an almost flat-fare scheme \citep{ben-akiva_discrete_1985}, rebound effects \citep{greening_energy_2000,hymel_induced_2010}, induced demand \citep{weis_induced_2009}, and global travel and activity patterns when paths costs are almost zero \citep[e.g.,][]{gonzalez_understanding_2008,alessandretti_evidence_2018}. The \NineEuroTicket{} is further a disruption to the currently fragmented fare structure in Germany: one single fare replaces fares in 60 different and not always contiguous transit districts. This simplification could also be a factor in attracting public transport users \citep{sharaby_impact_2012}. Consequently, it is not surprising that based on the drastic fare reduction and simplification the media reported that the people's sentiment toward public transportation has changed positively with the experience of the \NineEuroTicket{}.
At the policy level, it allows the analysis of to which extent in Germany an almost fare-free flat-fare scheme is an effective and efficient policy instrument to promote sustainable mobility, in particularly as it can be seen as an incentive to change from the car to public transport to reduce automobile externalities \citep{parry_automobile_2007,Santos2010}. However, as this policy does not address car dependence \citep{banister_sustainable_2008} and evidence suggests that a `\textit{coordinated package of mutually reinforcing transport and land-use policies that [...] (make) car use slower, less convenient, and more costly}' is required to reduce the share of car trips \citep{buehler_reducing_2016}, especially limiting parking supply \citep{loder_optimal_2022,shoup_high_1997,topp_parking_1993}, it can be expected that this policy cannot be the sole instrument to achieve sustainable mobility. From a policy analysis perspective it is interesting to observe the effect of public financial incentives (here reduction of ticket costs) in the transport sector, as one sort of policy instruments potentially initiating transformative change \citep{djeffal_role_2022}. While this sort of instruments played a rather limited role for the expansion of renewables, since regulatory and market instruments (like feed-in tariff system, \cite{hoppmann_compulsive_2014}) shaped energy transition in many countries \citep{barnea_policy_2022}, the role of public financial incentives in other fields, like the transport sector, is still unclear. 

The \NineEuroTicket{} can be considered to make Germany's local, regional, and urban public transport system almost fare-free as its price is very low and with a discount of more than 90~\% compared to existing travel passes, while neglecting the added value through its nationwide validity. In addition, the \NineEuroTicket{} is a nationwide flat-rate travel pass. For both aspects, precedent around the world exists. Fare-free public transport exists or existed in more than 100 locations \citep{keblowski_why_2020}. Fare-free public transport can be full or socially or temporally limited, e.g., to the elderly or to weekends. It is usually found in small urban areas with a population of less than 100~000, where recently larger areas have started fare-free public transport schemes as well. For example, Tallinn (Estonia) with a population of 400~000, made local public transport fare free in 2013. Here, a study reported that the increase in ridership that can be attributed to the fare innovation was 1.2~\% when controlling for other effects \citep{cats_public_2014}. Right amid in the beginning of the COVID-19 pandemic, three more locations started fare-free public transport. The entire country of Luxembourg \citep{carr_mobility_2020}, population of 630~000, but so far no report on the outcomes exists; Cascais (Portugal), population of 220~000, reported an increase in 10~\%, while the scheme is funded by parking fees \citep{kollinger_cascais_2022}; Augsburg (Germany), population of 300~000, defined a fare-free public transport zone in the city center \citep{andor_kostenloser_2021}, but so far no report on the outcomes exists. In an controlled experiment in Santiago (Chile), some workers got randomly assigned fare-free travel passes for two weeks \citep{bull_impact_2021}. The authors find that travel increases by 12~\%, while they find no evidence for mode or period substitution. The effect on public transport trips entirely explained by subway trips and by a residence location next to a subway station. In Templin (Germany), a similar pattern was observed with limited shift from the car, while a positive net benefit remains caused by a reduction in fatalities and casualties of pedestrians and cyclists \citep{storchmann_externalities_2003}. Generally, findings suggest that free-fare public transport schemes have the risk of generating additional travel demand, while do not necessarily encourage a shift from the car to public transport \citep{keblowski_why_2020,storchmann_externalities_2003}. It is further argued that such schemes are rarely implemented in a response to solve economic, sustainability, or socio-economic problems, which would be usually expected when implementing a transport policy, but rather did the institutions have different targets in mind \citep{carr_mobility_2020,keblowski_why_2020}. In a recent but pre-pandemic survey in Germany, \cite{andor_kostenloser_2021} reports an increase of about 230~\% in public transport usage, which serves as one reference for this study. Lately, in response to COVID-19, to promote sustainable transportation, and to improve air quality some transport operators in the US experiment with (partial) fare-free and heavily discounted schemes \citep{paget2023}.

At the beginning of public transport, fares were collected on a per-trip basis in cash, while the idea of a travel pass arose later, partly due to operating convenience. This introduction changed the product from a trip to access to a whole network. With the marginal cost for each trip becoming zero, its cost perception thus became similar to that for the car \citep{white_development_1984}. If travelers have a choice between pay-per-use and a flat-rate ticket, many choose the latter despite not reaching the break-even point. This flat-rate bias can be addressed mostly to an insurance and convenience effect \citep{wirtz_matthias_flatrate_2015}. It is noted that travel choices and economic outcomes for operators depend on the fare structure and distribution of travel demand \citep{carbajo_economics_1988}. Nationwide travel passes exist in many cities and even entire countries, e.g., in Switzerland with the well-known \textit{abonnement général} \citep{becker_modeling_2017} and since 2021 the \textit{KlimaTicket} in Austria. The latter travel pass builds on the idea of travel passes for 1~Euro per day or 365~Euro per year, which has seen much interest worldwide and has been implemented, e.g., in Vienna. It is reported that the volume of sold annual travel passes grew substantially, but that no direct significant increase in ridership or change in modal split resulted \citep{civity_management_consultants_gutachten_2021}. Nevertheless, Vienna continued to employ accompanying measures to restrict car traffic and parking that together with an increasing awareness for the high cost of parking, led to a decrease in car travel and increase in public transport use \citep{buehler_reducing_2016,knoflacher_wem_2019}.

The \NineEuroTicket{} can also be seen as a congestion management policy as it should attract people from cars, where the textbook rationale is that the subsidy for public transport results from road user charges \citep{small_economics_2007}. Nevertheless, optimality in fare subsidy depends primarily on the objective function, e.g., social cost \citep[e.g.,][]{tirachini_multimodal_2014} or user cost \citep[e.g.,][]{loder_optimal_2022}. \cite{basso_efficiency_2014} concluded that the substitutability between transit subsidy, like the \NineEuroTicket{} and congestion pricing, is large, but the contribution of transit subsidy could be severely diminished if other policies like congestion pricing are implemented first; otherwise,  the authors concluded that ``transit subsidies may indeed be a very good measure for reducing transport negative externalities and increase social welfare''.  

In this paper, we describe our study to observe this natural experiment in Germany with a focus on the Munich metropolitan region. The study comprises a three-wave survey conducted before, during, and after the validity period of the 9-Euro-Ticket in June, July, and August, and a smartphone-based travel diary with passive tracking to automatically measure individual travel behavior from May to September 2022 (see Figure \ref{fig:study_design}). The most relevant research questions of this transport policy experiment are: how did people change their travel behavior during the \NineEuroTicket{} intervention, who changed by how much, and how much do people value this ticket in the end. We present the recruiting and sample as well as present first findings on previously mentioned research questions.

This paper is organized as follows. Section \ref{sec:design} presents the study design. Then, Section \ref{sec:recruitement} reports on the recruitment of the sample as well as the participation during the study period; it also informs about the sample characteristics of our study. Thereafter, the results are presented in three sections: Section \ref{sec:sp_travel} presents the stated responses on changes in travel behavior throughout the study period; Section \ref{sec:app} reveals the changes in travel behavior collected through the smartphone; and Section \ref{sec:consumer} provides the consumer perspective on the \NineEuroTicket{} with the participants' attitudes towards the ticket and their willingness-to-pay for a successor ticket to the \NineEuroTicket{}. The paper ends with a discussion and conclusions of findings and their wider policy implications.

\section{Study design} \label{sec:design}
\input{02_studydesign.tex}

\section{Recruitment and participation} \label{sec:recruitement}
\input{03_recruitment.tex}

% -------------------------------------------------
%
\section{Stated travel behavior} \label{sec:sp_travel}
\input{04_survey.tex}

% -------------------------------------------------
\section{Smartphone-based travel diary} \label{sec:app}
\label{sec:ftm_travel_behavior}

\input{05_app.tex}

% -------------------------------------------------
%
\section{Consumer perspective on the 9-Euro-Ticket} \label{sec:consumer}
\input{06_consumer.tex}

% ----
\section{Discussion}

\citep[e.g.,][]{kramer_9-euro-ticket_2022} or using third-party mobile phone data \citep[e.g.,][]{gaus_9-euro-ticket_2023}

The changes in travel behavior during the period of the \NineEuroTicket{} are as expected: a shift towards public transport, but despite a very low price, not all car trips have been replaced by public transport (c.f. Figure \ref{fig:share_of_trips}). Some findings suggest that some study participants increased their public transport usage after the \NineEuroTicket{} intervention compared to the period before the ticket, but it remains to be seen how this effect develops over time. With the introduction of a successor ticket in May 2023 priced at 49\,Euro, it will be interesting to how the different pricing points and the non-temporary nature of the successor ticket make a difference in travel behavior compared to the \NineEuroTicket{} period.

The responses of our three-wave panel survey are in alignment with the general perception of the effects of the \NineEuroTicket{} in Germany, arguably due to relying on a professional panel agency to ensure a representative sample. Nevertheless, the rates of \NineEuroTicket{} ownership in the Munich-based panel at around 90\,\% seem rather high. This could result from a sample selection bias as the ``Mobilit\"{a}t.Leben'' study was marketed in the media as the \NineEuroTicket{} study, but given the good accessibility by public transport in the Munich metropolitan region, it can not be ruled out that these ownership levels are representative for the region. Unfortunately, no local data exists to validate this hypothesis. 

This study bears limitations. The short time period between the announcement of the \NineEuroTicket{} and its start in June 2022 did not suffice to recruit a representative panel across the Munich metropolitan area. Despite relying on a professional panel agency for the three-wave panel survey to partially account for this in the survey, the sample bias in our tracking sample persists. Nevertheless, all relevant parts of the population are part of our sample and thus sample weights that account, e.g., for income and education, can be derived to reduce the bias. The short time period also limited the observation time before the start of the \NineEuroTicket{} with half of the sample having recorded no trip at all before the ticket's introduction. Thus, appropriate methodological techniques must be selected when investigating the impact of the \NineEuroTicket{} on travel behavior like mode choice and induced demand. 

Generally, the used innovative travel survey method of combining a panel survey with a smartphone-based travel diary has been proven successful. This study faces similar participation issues, e.g., self-selection bias, drop outs, and missing validation of trips and activities, as other studies have reported \citep[e.g.,][]{allstrom_smartphone_2017,lugtig_nonresponse_2022}, that can limit the validity of results. Nevertheless, with its focus, design and length, the study ``Mobilit\"{a}t.Leben'' will make contributions to travel survey methods and thus allow to improve future travel surveys using smartphone-based travel diaries together with panel surveys. 

\section{Conclusions}

In this paper, we presented the first empirical findings of our study ``Mobilit\"{a}t.Leben'' that we established to observe the natural and forced experiment of the \NineEuroTicket{} in the summer months of 2022 in Germany. We showed the stated responses collected through a three-wave panel survey on the changes in travel behavior with focus on car and public transport usage. We corroborated these findings with the data collected through our smartphone-based travel diary with passive waypoint tracking. In our tracking sample we found a modal shift towards public transport from the car of around five percentage points of the average daily travel distance during the \NineEuroTicket{} period, which is similar to findings from another study \citep{gaus_9-euro-ticket_2023}. We also found indications in the collected data that suggest that some participants increased their public transport usage after the \NineEuroTicket{} intervention compared to the period before. In other words, some participants might be activated to increase public transport usage through the \NineEuroTicket{}. We also investigated the willingness-to-pay for a successor ticket to the \NineEuroTicket{} during and after the \NineEuroTicket{}. When considering that the successor ticket started on May 1, 2023, priced at 49~Euro, we found that it will impact travel pass ownership: in our nationwide panel ownership will increase only slightly, while in our Munich-based panel ownership levels could grow by up to 50\,\%. We further found that if public transport operators provide an attractive local travel pass in addition to the successor ticket, ownership shares for travel passes could rise even further.

Future research will explore in depth the behavioral dimensions of the natural and forced experiment. Particular focus will be on investigating the impact of the \NineEuroTicket{} on mode choice and associated changes in the valuation of travel time as well as on analyzing induced demand effects. As it is known that crowding increases the value of travel time savings \citep[e.g.,][]{li_crowding_2011,prudhomme_public_2012,wardman_passengers_2015}, future research will also explore the impact of crowding. Given the special spatial feature of the \NineEuroTicket{} compared to the existing travel passes with limited spatial validity in Germany, particular focus will be given to travel behavior across administrative boundaries. With the desire of the \NineEuroTicket{} to be a strong policy instrument for reducing carbon emissions in the transportation sector, future research will also concentrate on estimating the nationwide carbon emission savings from the \NineEuroTicket{} policy in relation to the simultaneous fuel tax cut. Future research will also analyze the primary nature of the \NineEuroTicket{} of being a relief to the 2022 cost-of-living crisis and to what extend this subsidy policy had on the redistribution of income \citep{spence_nonlinear_1977}. Last, the travel survey method used in our study will be analyzed and compared to other approaches too. From a political science perspective it will be interesting to analyze the long-term impact of the 49-Euro-Ticket on the policy mix in the transport policy field, and to what extent demands in the population for a fundamental transformation of the transport sector (towards more public transport) will be promoted or inhibited.

In closing, before the start of the \NineEuroTicket{} in June 2022, many expected overcrowded trains. In the end, this did not occur on a regular basis everywhere in the country, but only a few trains to touristic locations, e.g., from Berlin to the Baltic Sea or from Munich to Lake Tegernsee. Additionally, congestion levels did drop, but congestion did not vanish. The \NineEuroTicket{} experiment shows that the effects of almost fare-free public transport in Germany are not substantially different from other experiences \citep[e.g.,][]{cats_public_2014}, allowing us to make the following key policy implication: the desire that attractive pricing alone suffices to incentive sustainable travel behavior and lift congestion did not materialize; other or additional policy instruments are required to meet our societies' sustainability goals.  

\section*{Acknowledgement}

Allister Loder acknowledges funding by the Bavarian State Ministry of Science and the Arts in the framework of the bidt Graduate Center for Postdocs. Fabienne Cantner and Felix Gotzler acknowledge funding by the Munich Data Science Institute (MDSI) within the scope of its Seed Fund scheme. The authors would like to thank the TUM Think Tank at the Munich School of Politics and Public Policy led by Urs Gasser for their financial and organizational support and the TUM Board of Management for personally supporting the genesis of the project. The authors thank the company MOTIONTAG for their efforts in producing the app at unprecedented speed. Further, the authors would like thank everyone who supported us in recruiting participants, especially Oliver May-Beckmann and Ulrich Meyer from M Cube and TUM, respectively.

%\section*{Author contributions}

%Allister Loder: Conceptualization, Data Curation, Formal analysis, Investigation, Writing - Original Draft.
%Fabienne Cantner: Conceptualization, Data Curation, Formal analysis, Investigation, Writing - Original Draft, Writing - Original Draft.
%Lennart Adenaw: Conceptualization, Software, Data Curation, Formal analysis, Investigation, Writing - Original Draft
%Nico Nachtigall: Software, Formal analysis, Writing - Original Draft, Visualization, Data Curation.
%David Ziegler: Software, Formal analysis, Writing - Original Draft, Visualization, Data Curation.
%Felix Gotzler: Software, Formal analysis, Writing - Original Draft, Visualization, Data Curation.
%Markus B. Siewert: Project administration, Funding acquisition, Supervision, Writing - Original Draft.
%Stefan Wurster: Supervision, Writing - Original Draft.
%Sebastian Goerg: Conceptualization, Supervision, Writing - Original Draft.
%Markus Lienkamp: Resources, Supervision, Writing - Original Draft.
%Klaus Bogenberger: Conceptualization, Resources, Supervision, Writing - Original Draft.

\bibliography{references,extra_bib}

\end{document}

%% file: 02_studydesign.tex
\begin{figure}
    \centering
    \includegraphics[width=\textwidth]{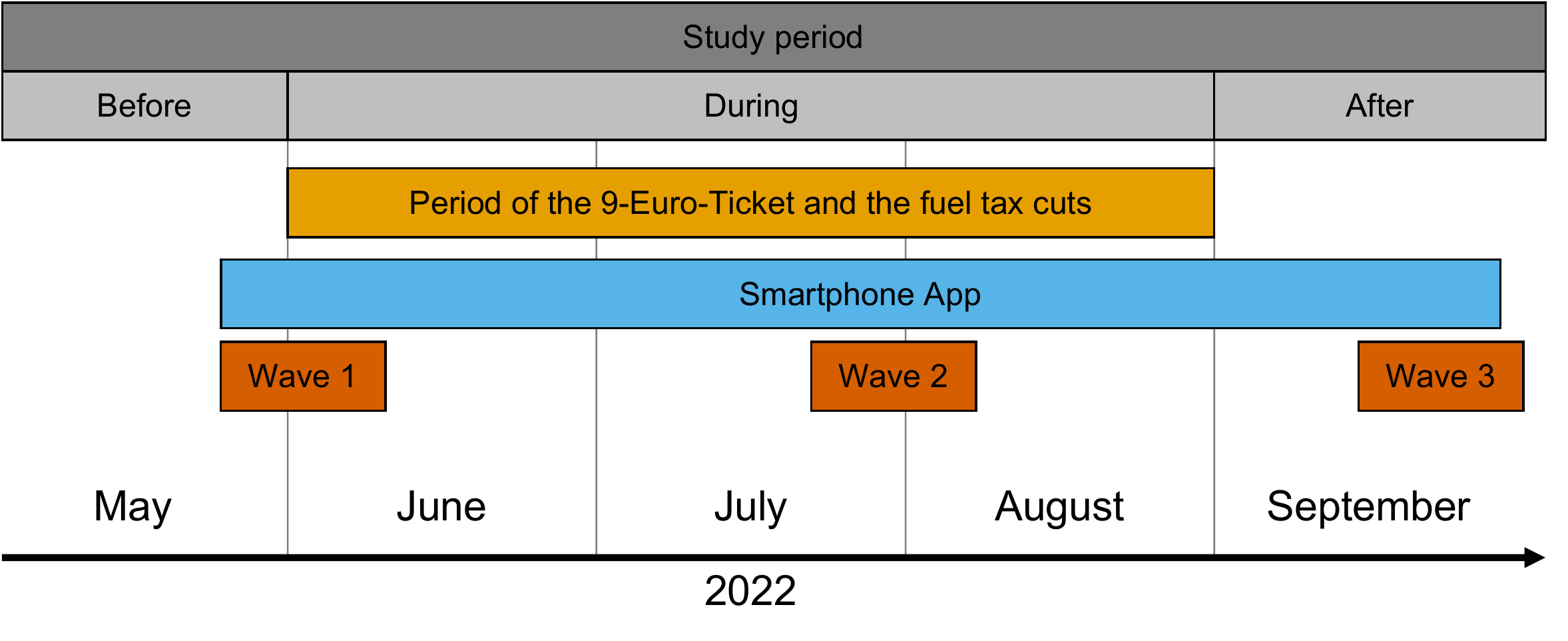}
    \caption{Design of the ``Mobilit\"{a}t.Leben'' study during summer 2022. }
    \label{fig:study_design}
\end{figure}

We conceived the name ``Mobilit\"{a}t.Leben'' for the study (in English ``mobility.life'') to characterize that our study intends to observe the impact of the cost-of-living crisis on mobility and the daily activities. The overall design of the ``Mobilit\"{a}t.Leben''  study is shown in Figure \ref{fig:study_design}. It follows the rhythm of the behavioral intervention of the \NineEuroTicket{} and the fuel tax cut: it comprises a period \textit{before} the intervention, one \textit{during} the intervention, and a period \textit{after} the intervention. In the three-wave panel study, a questionnaire is distributed to participants in each of the three periods. The structure and content of the questionnaires is detailed in Section \ref{sec:survey}, while the survey's codebook will be made available upon publication.
%\footnote{\url{https://www.hfp.tum.de/hfp/tum-think-tank/mobilitaet-leben/veroeffentlichungen/}}
The travel behavior of participants is recorded before, during and after the period of the \NineEuroTicket{} and the fuel tax cut using an application for a smartphone-based travel diary with passive tracking. The smartphone application is described in Section \ref{sec:app}. 

\subsection{Survey} \label{sec:survey}

The three-wave panel survey collects information of participants on their demographics (e.g., age, sex, employment status, education level), socio-economic status (e.g., household income), place and kind of residence (e.g., home location, heating system), personality traits (scales of the Global Preference Survey \citep{falk2018global, falk2022preference}, as well as social and political attitudes (e.g., political left-right positioning, support of the \NineEuroTicket{} policy, climate change awareness).  Regarding mobility tool ownership, participants report on household car ownership, season-ticket ownership before and after the \NineEuroTicket{} as well as the intention to buy and having the \NineEuroTicket{} in June, July and August. Given the cost-of-living crisis, the three questionnaires also feature questions on the financial situation of participants: how the household deals with the recent price increases, whether the household can save income or has to draw from savings, and whether the household is doing without consumption or investments to compensate for the price increases. In addition, the questionnaires include questions on energy consumption behavior and whether measures have been taken to conserve energy.
%\klaus{später kommt mal nation-wide panel und Munich panel, das ist hier nicht beschrieben ... }

We measure the changes in travel behavior during the experiment in three different ways. First participants were asked in each wave to state their current travel behavior using mode use frequencies following the Mobilität-in-Deutschland (MiD) scale \citep{mid2017}, i.e., how frequent do you use the car, public transport, bicycle etc. per week. Second, in the second and third wave participants were asked  to state qualitatively whether they use public transport and the car more or less often compared to the previous wave, i.e., comparing travel behavior during and before, after and during, and after and before. Third, participants were asked in the second and third wave to state how many more or less out-of-home activities they currently perform using the car and public transport.

\begin{figure}
     \centering
     \begin{subfigure}[b]{0.55\textwidth}
         \centering
         \includegraphics[width=\textwidth]{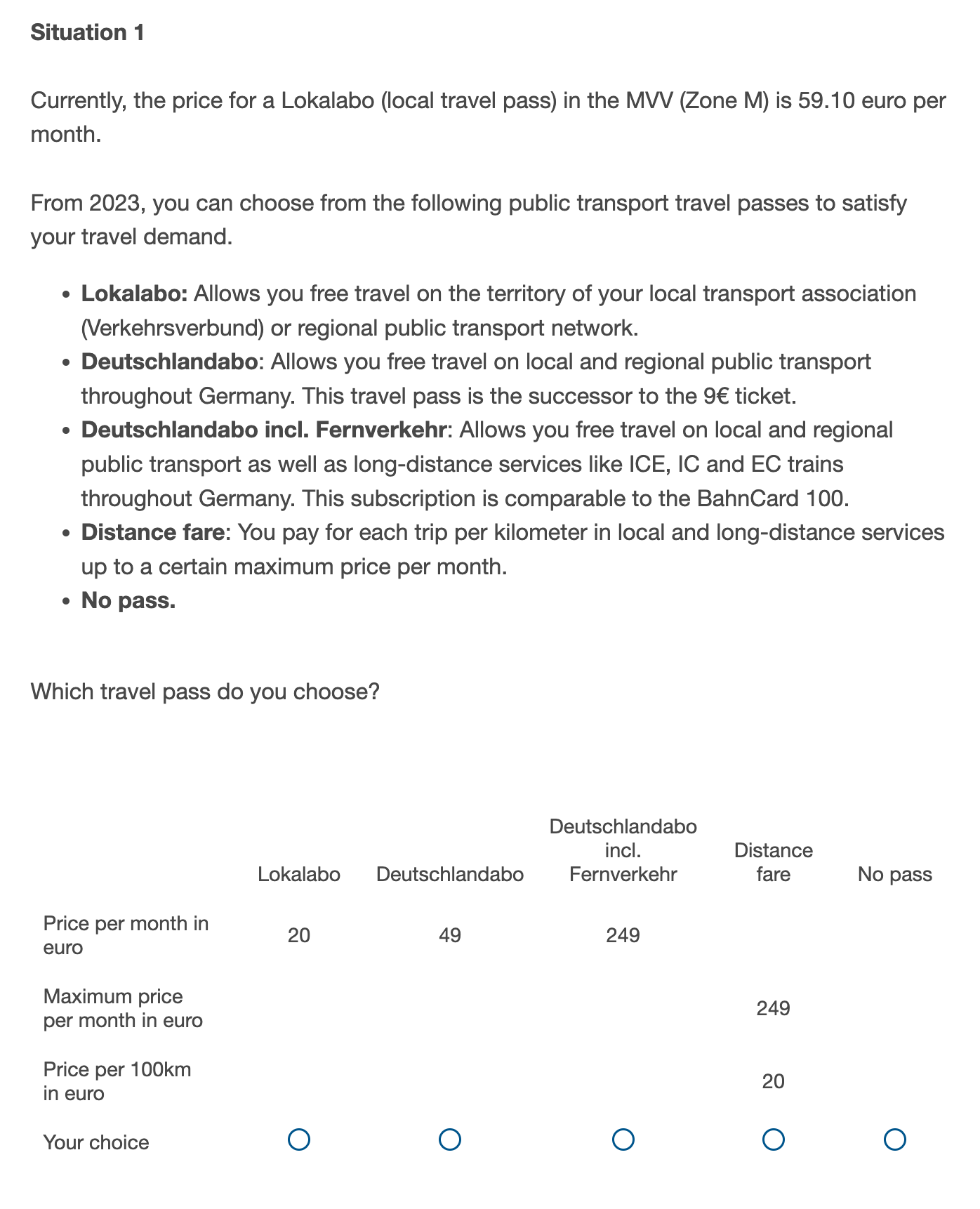}
         \caption{Screenshot of the stated-preference experiment in the third questionnaire.}
         \label{fig:sp_experiment}
     \end{subfigure}
     \hfill
     \begin{subfigure}[b]{0.3\textwidth}
         \centering
         \includegraphics[width=\textwidth]{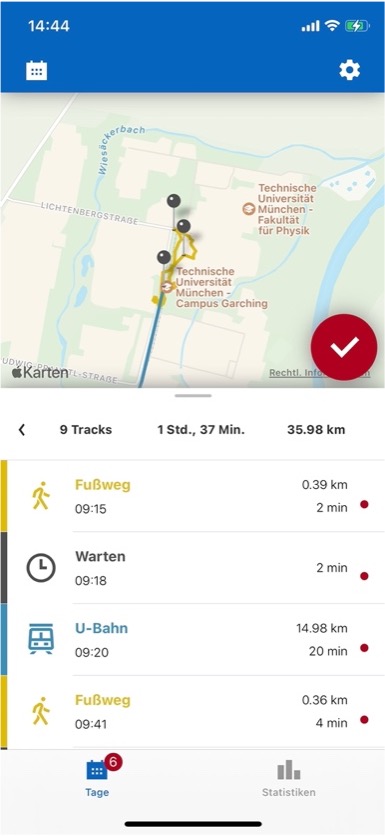}
         \caption{Screenshot of the smartphone-based travel diary with passive tracking.}
         \label{fig:app}
     \end{subfigure}
     \hfill
        \caption{Screenshots from the study ``Mobilit\"{a}t.Leben''. (a) from the stated-preference experiment on the successor ticket; (b) from the study's smartphone application.}
        \label{fig:exp_tools}
\end{figure}

As the \NineEuroTicket{} was a fix-term transport policy, the question of a successor to the \NineEuroTicket{} was very prominent in the public debate. Thus, we included questions on the interest to buy a successor ticket and its maximum willingness-to-pay in the second and third questionnaire. Further, to better mimic the decision-making process to buy a successor ticket, we included a stated-preference experiment in the third questionnaire. Figure \ref{fig:sp_experiment} shows a screenshot of one choice situation. In this experiment, participants had the choice to select either a local travel pass with regional validity (Lokalabo), a successor to the \NineEuroTicket{} (Deutschlandabo),  a successor to the \NineEuroTicket{} that includes all long-distance services (Deutschlandabo incl. Fernverkehr), a distance fare or no travel pass (outside option). A distance fare was included to see whether participants have at least some interest in a public transport product. In the experiment, only the price attribute of each alternative is varied: we set the price levels for the Lokalabo to 19 and 29~Euro; for the Deutschlandabo to 49, 59, 69, and 99~Euro; for the Deutschlandabo including long-distance services to 249 and 349~Euro, and the price levels for the distance fare to 10 and 20~Euro per 100~km. Deploying a full-factorial design leading to 32 scenarios, each participant were shown eight choice situations. 

\subsection{Smartphone-based travel diary} \label{sec:travel_diary}

For our study, a travel diary smartphone application, called "Mobilit\"at.Leben", has been developed. Such an application has already been used in other travel behavior research projects, e.g., in \cite{molloy_mobis_2022}. Figure \ref{fig:app} shows a screenshot of the developed app. Installed on iOS and Android devices, it automatically collects users' way-points, reports them to a backend server where the chosen transport mode is inferred using machine learning by a third-party provider. This data is returned to the smartphone application and users can approve or edit the mode. The smartphone application also records the time between trips as activities. The smartphone application suggests in some cases activity types, e.g. home, shopping, but in most cases users have to enter the activity type manually. However, the smartphone application learns these activities, reducing in the best case the required users input to complete the travel diary. There is no need for further interaction with the app. Participants can pause the data collection at any time.

%% file: 03_recruitment.tex
\subsection{Recruitment}

The sample for the ``Mobilit\"at.Leben'' study has been recruited following a two-pronged strategy. First, a media campaign in Munich started on May 23, 2022, resulting in 1,342 individuals having registered for the study. A compensation of 30~Euro was offered for participation when using the smartphone travel diary until September 2022 as well as completing all three questionnaires. Individuals who registered for the study had the choice between participating in the smartphone-based travel diary and the survey or only in the survey. Second, a professional panel agency was tasked with recruiting participants for the three-wave panel survey. These participants, however, were not using the smartphone-based travel diary. Individuals were offered a compensation for the participation directly through the panel agency. Here, 919 individuals have been successfully recruited. In total, 2,261 individuals started the ``Mobilit\"at.Leben'' study. In the following, we refer to the \textit{Tracking} sample as those participants from the Munich-based sample who use the smartphone-based travel diary and fill out the questionnaires and we refer to the \textit{Survey-only} sample as those participants from the nation-wide and Munich-based sample who do not participate in the smartphone-based travel diary, but fill out the questionnaires. 

It is important to mention that the 1,342 participants recruited through the media campaign are primarily located in the Munich metropolitan area. Consequently, we denote this part of our sample as the \textit{Munich-based panel}. Contrary, the  919 participants recruited through the panel agency are from all over Germany. Consequently, we denote this part of our sample as the \textit{nation-wide panel}. Generally, by asking the panel agency to recruit a representative sample, the \textit{nation-wide panel} can be considered more representative for the entire nation than the \textit{Munich-based panel}.

\subsection{Outlier removal}

To ensure the best possible data quality, we have taken several steps. First, at the beginning of all surveys we appeal to participants to answer the questions carefully and honestly. We also emphasize that they will only receive the total payment if the submit a fully completed survey. Second, we track the time spend on each of the surveys. For our analysis, we exclude respondents who completed the survey too fast. That is, we drop participants in the bottom 1\,\% of the survey time distribution. Finally, we exclude participants who gave inconsistent responses about purchasing the \NineEuroTicket{} across the three different survey periods. Exclusion of these respondents does not impact our results. In total, these conditions apply to 117 individuals who are subsequently removed from the further analysis.

Thus, out of the 2,261 registered participants, 2,144 are further considered in this analysis (94.8\%).  Table \ref{tab:survey_participation} shows the considered sample the participation in the three waves.  Here, completion means that all three questionnaires of the three-wave panel survey were completed. This results in 1,402 complete responses for the survey. Out of these, 637 also completed the smartphone-based travel diary with passive tracking by successfully providing at least one week of travel behavior in June, July, August and September.

%\klaus{Was ist mit outlier removal bei den app-tracking daten}

\subsection{Participation}

\begin{figure}
    \centering
    \includegraphics[width=14cm]{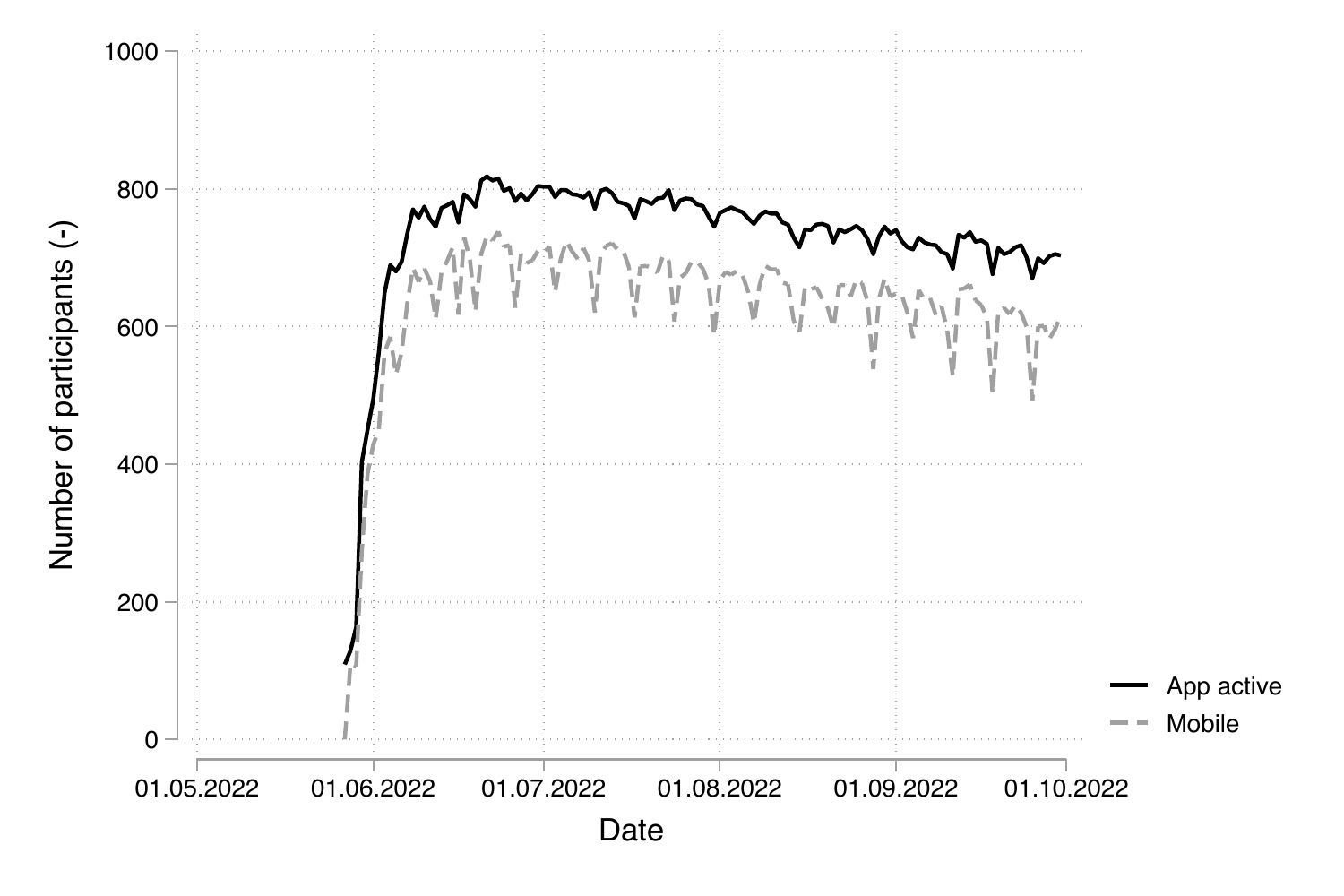}
    \caption{Study participation in the smartphone application Mobilit\"{a}t.Leben: the number of smartphones actively tracking and the number of mobile users per day.}
    \label{fig:app_heartbeat}
\end{figure}

The study participation is summarized in Table \ref{tab:survey_participation} and the smartphone-based travel diary usage is shown in Figure \ref{fig:app_heartbeat}. Table \ref{tab:survey_participation} provides information on the number of recruited, the completion of each questionnaire as well as the completion of all three questionnaires for the entire sample, the Munich-based panel, and the nation-wide panel. The Munich-based panel is further separated by participating in the smartphone-based travel diary (tracking sub-sample) or only in the survey (survey-only sub-sample). All participants from the Munich-based panel that use the smartphone-based travel diary are considered in Section \ref{sec:ftm_travel_behavior}, not only those who successfully completed the study.

\begin{figure}
    \centering
    \includegraphics[width=14cm]{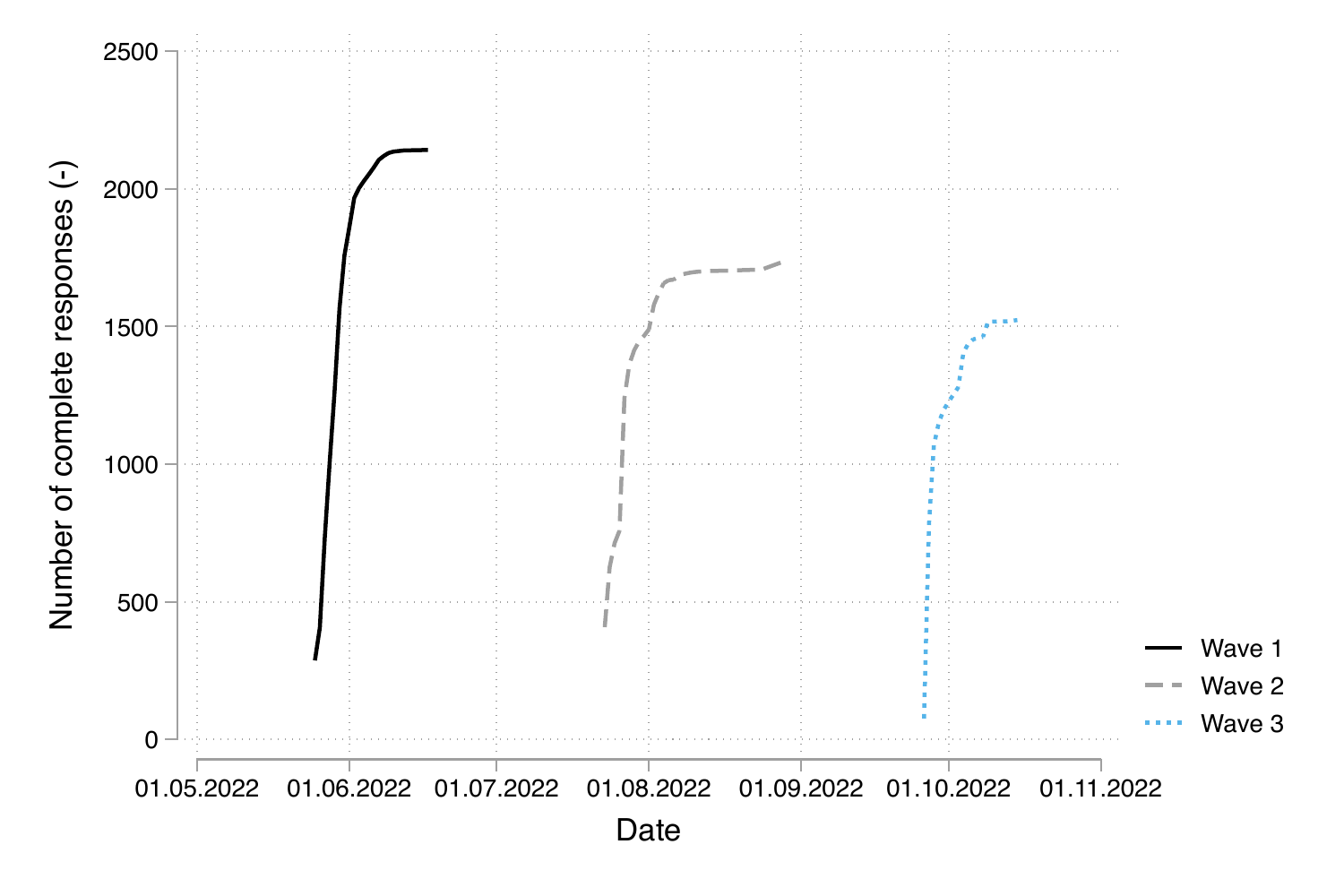}
    \caption{Time series of the participation in the three-wave survey of the entire sample.}
    \label{fig:survey_part}
\end{figure}

It can be seen in Table \ref{tab:survey_participation} that the drop out rate is higher in the Munich-based survey-only panel where no incentive was provided compared to the Munich-based panel that uses the smartphone application (30~Euro incentive) and the nation-wide survey panel (incentive through panel agency). At the end of the study period in September, the study has not been completed by all participants. In total, 1,402 participants completed all three questionnaires. In the Munich-based panel that uses the smartphone application, around 80\% who started the tracking completed all three questionnaires and around 70.0\% completed the tracking and all three questionnaires. In the Munich-based survey-only panel, only around 40\% completed all three surveys and in the nation-wide panel around 62\%. Figure \ref{fig:survey_part} shows the completion of the three questionnaires over time. It can be seen that most participants completed the survey rather quickly after receiving the invitation.

\begin{table}[]
    \centering
    \begin{tabularx}{\textwidth}{X|rr|rr|rr|rr}
    \toprule
     & \multicolumn{2}{c}{\multirow{2}{*}{Total}} & \multicolumn{4}{c}{Munich-based panel} & \multicolumn{2}{c}{\multirow{2}{*}{Nation-wide panel}} \\ \cmidrule{4-7}
     & \multicolumn{2}{c}{} & \multicolumn{2}{c}{Tracking} & \multicolumn{2}{c}{Survey only} & \multicolumn{2}{c}{} \\
     \midrule
Registered & 2,144 & 100.0\% & 871 & 100.0\% & 401 & 100.0\% & 872 & 100.0\% \\
Completed 1st wave & 2,025 & 94.4\% & 855 & 98.2\% & 298 & 74.3\% & 872 & 100.0\% \\
Completed 2nd wave & 1,631 & 76.1\% & 765 & 87.8\% & 217 & 54.1\% & 649 & 74.4\% \\
Completed 3rd wave & 1,436 & 67.0\% & 713 & 81.9\% & 180 & 44.9\% & 543 & 62.3\% \\
Completed all three waves & 1,402 & 65.4\% & 695 & 79.8\% & 164 & 40.9\% & 543 & 62.3\%\\
\bottomrule
\end{tabularx}
    \caption{Partition in the three-wave panel survey of the entire sample and all relevant sub-samples.}
    \label{tab:survey_participation}
\end{table}

Considering the smartphone application usage shown in Figure \ref{fig:app_heartbeat}, it can be seen that the app's activation across the sample happened during the introductory phase of the \NineEuroTicket{} in the early days of June. Note that an earlier activation of the smartphone app was barely possible as the natural experiment was announced at short notice by the government, leaving little time for recruitment weeks before the introduction of the \NineEuroTicket{}. Overall, as stated in Table \ref{tab:survey_participation}, 920 participants successfully activated the application on their smartphone during our study; smartphone-based travel diary usage peaked in late June at around 820 users tracking in parallel. Here, some users already deleted the application, e.g., due to high energy consumption, or switch it off on some days, e.g., when leaving Germany. The number of active smartphones declined steadily towards around 700 users at the end of our study, while the observed wave pattern indicates that a fraction of the sample is frequently switching the tracking on and off. Figure \ref{fig:app_heartbeat} also features the number of mobile users on each day. Here, the typical weekly pattern is observed with more mobile people during the week compared to weekends. 

\subsection{Study sample}
\label{sec:final_sample}

\begin{figure}
    \centering
    \includegraphics[width=\textwidth]{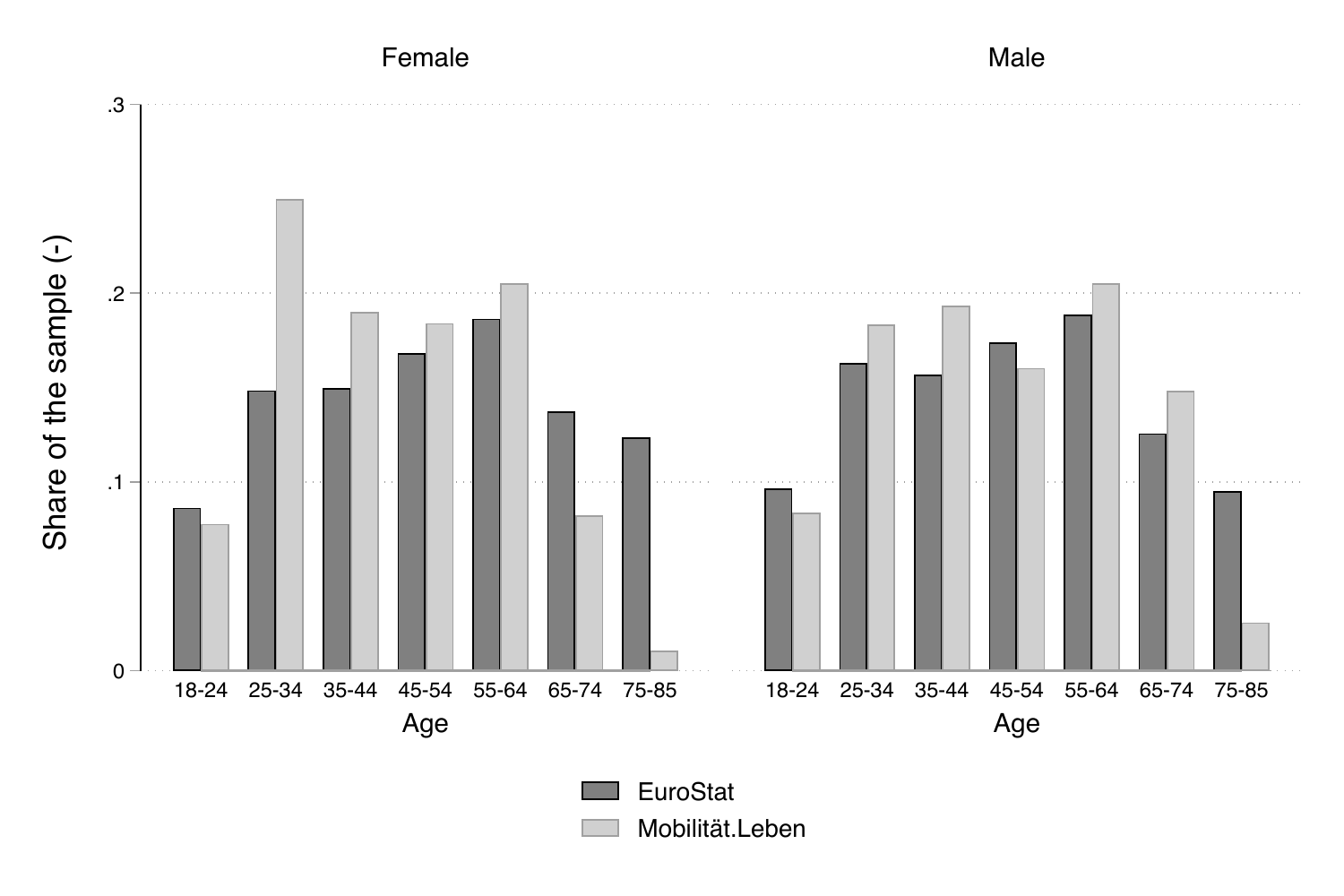}
    \caption{Comparing the Mobilit\"{a}t.Leben sample to EuroStat's population data for Germany for both sexes and age categories.}
    \label{fig:age_distribution}
\end{figure}

The characteristics of the final sample for the survey and the tracking is shown in Table \ref{tab:sample_info}; the table further provides information on the sample that started the tracking. We compare the socio-demographic composition of final sample and the sub-sample that completed also the smartphone-based travel diary with passive tracking to the 2017 German national household travel survey, Mobilit\"{a}t-in-Deutschland (MiD), \citep{mid2017}. We conclude that both samples have slightly more men than women compared to the MiD. Considering age groups, the age distribution of both samples differ only substantially in the age groups 25-34 and $>$ 75 from the MiD. There are two reasons. First, the own recruiting via a media campaign originated at a university, where the recruitment of more young people seems reasonable. Second, the group $>$ 75 is underrepresented likely as the study had only digital interfaces (survey and smartphone), where this age group is less familiar with; further, the numbers cannot really be compared in this group as MiD only reports the total of surveyed people in the group $> 75$ without any further intervals. Thus, to better compare the age distribution, we compare our completed survey sample to official age distribution by gender in \ref{fig:age_distribution} using data from EuroStat \citep{eurostat2023}. Here, we conclude that our sample has similar distribution to the general population with notable exceptions: in the group 75-85 we have too few and the share of female participants in the age group 25-34.

\include{TABLES/sample.tex}

Regarding the household size, both our samples have more single households compared to the MiD that can be likely attributed, again, to the partial recruitment in an university environment with many individuals not yet having started a family.

The comparison of net household income in our sample to the MiD faces the methodological challenge that income is not directly reported in the national household travel survey, but rather a derived composite measure of a household's economic status that combines net household income and weighted sum of household members. In Table \ref{tab:sample_info} we approximated this measure for our sample; we conclude that both our samples have a substantial oversampling of households with a high economic status. This could be a consequence from the Munich metropolitan area being focus region of our study, which is characterized by having wages and income above the nation's averages. 

The focus on Munich is then further reflected in the distribution of our sample across the Germany's statistical spatial types, car ownership and travel pass ownership in May 2022. More individuals living in the metropolis and less in the countryside, resulting in less car ownership and higher travel pass ownership levels compared to the MiD.

\subsection{Sample weighting} \label{sec:sample_weighting}

The recruited sample is not a random sample of the German population. Although it can be seen in Figure \ref{fig:age_distribution} that the complete survey sample does exhibit some similarity with Germany's general age distribution by gender, there are differences. Therefore, we calculate probability weights for the complete sample based on age and gender in the categories shown in Figure \ref{fig:age_distribution}. The obtain weights are reasonable in the range of 0.8 to 2 for all age groups below 76-85, while the latter age group has values from 5-12, reflecting the under representation of elderly in the sample.

%% file: TABLES/sample.tex
\begin{sidewaystable}
\small
\caption{Comparison of sample characteristics to the \gls{mid} travel survey. Missing totals to one hundred percent in each category are missing values. The column ``Tracking started'' also contains all observations that  started, including those eliminated as named in Section \ref{sec:final_sample}.}
\label{tab:sample_info}
\begin{tabularx}{\textwidth}{lX|r|rr|rr|rr}
\toprule
 &  & \multicolumn{1}{c}{MiD} & \multicolumn{6}{c}{Mobilität.Leben final sample} \\ \cmidrule{4-9}
 &  & \multicolumn{1}{c}{2017} & \multicolumn{2}{c}{Survey completed}  & \multicolumn{2}{c}{Tracking completed} & \multicolumn{2}{c}{Tracking started} \\
 &  & \multicolumn{1}{c}{\%} & \multicolumn{1}{c}{N} & \multicolumn{1}{c}{\%} & \multicolumn{1}{c}{N} & \multicolumn{1}{c}{\%} & \multicolumn{1}{c}{N} & \multicolumn{1}{c}{\%} \\
 & Total & - & 1,402 & \multicolumn{1}{r}{100.00\,\%} & \multicolumn{1}{c}{637} & \multicolumn{1}{c}{100.00\,\%} & \multicolumn{1}{c}{920} & \multicolumn{1}{c}{100.00\,\%} \\
 \midrule
Gender & Male & 48.86\,\% & 779 & 55.56\,\% & 339 & 53.22\,\% & 491 & 54.37~\% \\
 & Female & 51.12\,\% & 619 & 44.15\,\% & 295 & 46.31\,\% & 425 & 46.20~\%  \\
 & Diverse & 0.03\,\% & 4 & 0.28\,\% & 3  &  0.47\,\% & 4 & 0.43~\% \\
 \midrule
Age & 18-24 & 9.15\,\% & 103 & 7.35\,\% & 59  &  9.26\,\% & 97 & 10.54~\% \\
 & 25-34 & 15.26\,\% & 299 & 21.33\,\% & 171  &  26.84\,\% & 262 & 28.48~\% \\
 & 35-44 & 14.29\,\% & 263 & 18.76\,\% & 118 &  18.52\,\% & 166 & 18.04~\% \\
 & 45-54 & 19.60\,\% & 239 & 17.05\,\% & 99 & 15.54\,\% & 130 & 14.13~\% \\
 & 55-64 & 16.32\,\% & 292 & 20.83\,\% & 108 &  16.95\,\%  & 160 & 17.39~\%\\
 & 65-74 & 12.01\,\% & 173 & 12.34\,\% & 64 &  10.05\,\% & 84 & 1.52~\% \\
 & $>$ 75  & 13.05\,\% & 27 & 1.93\,\% & 12  &  1.88\,\% & 14 & 0.76~\% \\
 \midrule
Household size  & 1 & 19.72\,\%  & 381 & 27.18\,\% & 175  & 27.47\,\% & 243 & 26.41~\%  \\
                & 2 & 38.49\,\% & 593 & 42.30\,\% & 260  & 40.82\,\% & 381 & 41.41~\% \\
                & 3 & 19.36\,\% & 204 & 14.55\,\% & 99 & 15.54\,\% & 131 & 14.24~\%  \\
                & 4 or more & 22.42\,\%  & 224 & 15.98\,\% & 103 & 16.17\,\% & 165 & 17.93~\%  \\
 \midrule
 Net household income & $<$ 1,500~Euro/month & - & 177 & 12.62\,\% & 73 & 11.46\,\% & 114 & 12.39~\% \\
        & 1,500 to 2,499~Euro/month & - & 256 & 18.26\,\% & 82 & 12.87\,\% & 123 & 13.37~\% \\
  & 2,500 to 3,999~Euro/month & - & 394 & 28.10\,\% & 180 & 28.26\,\% & 245 & 26.63~\% \\
  & $>$ 4,000~Euro/month & - & 547 & 39.02\,\% & 283 & 44.43\,\% & 409 & 44.46~\% \\
 \midrule
Economic status & Very low      & 7.33\,\% & 65 & 4.63\,\% & 37 & 5.81\,\% & 62 & 6.74~\% \\
                & Low           & 13.12\,\% & 171 & 12.22\,\% & 57 & 8.79\,\% & 76 & 8.26~\% \\
                & Average       & 43.64\,\% & 322 & 22.82\,\% & 112 & 17.58\,\% & 164  & 17.83~\% \\
                & High          & 29.94\,\% & 513 & 36.45\,\% & 251 & 39.40\,\% & 341  & 37.07~\% \\
                & Very high     & 5.97\,\% & 303 & 21.70\,\% & 162 & 25.43\,\% & 234 & 25.43~\% \\
 \midrule
Regional statistical spatial   type & Urban: Metropolis & 18.14\,\% & 653 & 46.58\,\% & 407 & 63.89\,\%  & 600 & 64.47~\% \\
                        & Urban: Regiopolis, large city & 14.61\,\% & 112 & 7.99\,\% & 20  & 3.14\,\% & 29 & 3.15~\% \\
 & Urban: Medium-sized city, urbanized area & 24.74\,\% & 275 & 19.61\,\% & 113 & 17.74\,\% & 154 & 16.74~\% \\
 & Urban: Small-town area, village area & 6.10\,\% & 71 & 5.06\,\% & 31 & 4.87\,\% & 42 & 4.57~\%  \\
 & Rural: Central city & 5.97\,\% & 47 & 3.35\,\% & 9 & 1.41\,\% & 13 & 1.41~\%  \\  
 & Rural: Medium-sized city, urbanized area & 14.32\,\% & 116 & 8.27\,\% & 21 & 3.30\,\% & 34 & 3.70~\% \\
 & Rural: Small-town area, village area & 16.12\,\% & 115 & 8.20\,\% & 29 & 4.55\,\%  & 39 & 4.24~\% \\
 \midrule
Car ownership & None & 14.71\,\% & 351 & 25.05\,\% &  210 & 32.97\,\% & 256 & 27.83~\%  \\
 & 1 cars & 47.26\,\% & 667 & 47.57\,\% & 291 & 45.68\,\%  & 364 & 39.57~\% \\
 & 2 cars and more & 38.02\,\%  & 384 & 27.39\,\% & 136  & 21.35\,\%  & 300 & 32.61~\% \\
 \midrule
Travel pass   ownership May 2022 & Yes & 14.66\,\% & 467 & 33.31\,\% & 283 & 44.43\,\% & 410 & 44.57~\% \\
 & No & 85.15\,\% & 935 & 66.69\,\% & 354 & 55.57\,\% & 490 & 53.26~\% \\
 \bottomrule
\end{tabularx}
\end{sidewaystable}

%% file: 04_survey.tex
A three-wave panel survey allows us to investigate the travel behavior changes likely caused by the intervention of the \NineEuroTicket{}. We investigate the travel behavior changes by considering the weekly mode use frequencies in each of the three periods; to corroborate the findings we then analyze questions of currently more or less mode use compared to the previous period. Last, we present the self-reported substitution of car trips for public transport.

\begin{figure}
    \centering
    \includegraphics[width=\textwidth]{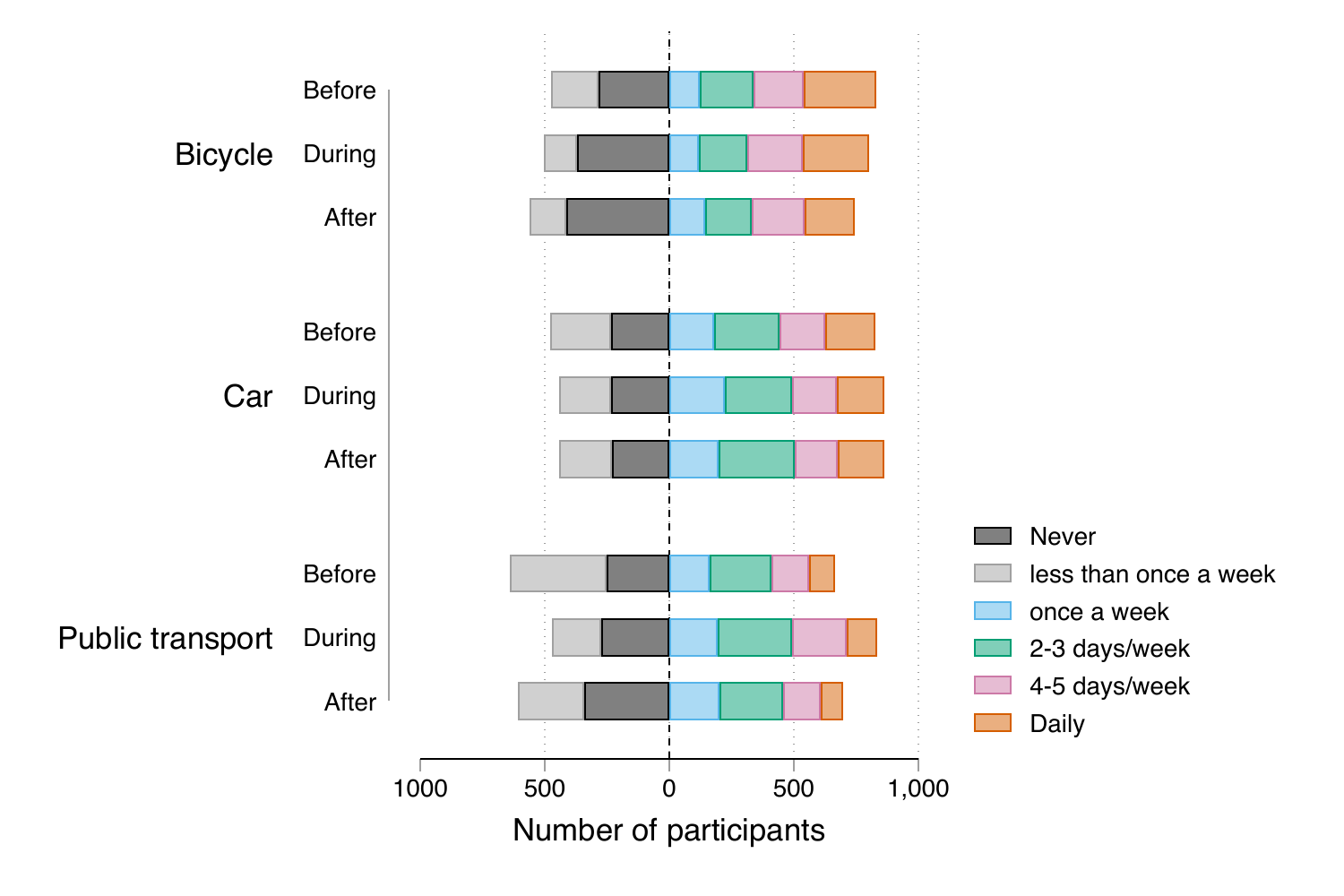}
    \caption{Mode use frequencies before, during, and after the \NineEuroTicket{} intervention.}
    \label{fig:trip_freq}
\end{figure}

The weekly mode use frequencies are reported on a six-level scale: never, less than once a week, once a week, 2-3 days per week, 4-5 days per week, and almost daily. Figure \ref{fig:trip_freq} shows the responses of the complete survey sample (N=1,402) for bicycle, car, and public transport use for the periods of before, during, and after as defined in Figure \ref{fig:study_design}. It can be clearly seen that public transport use increased substantially the during period, while car use did not see similar reductions. Table \ref{tab:trip_freq_changes} shows how the sample distributes across the reported changes in mode use frequencies from before to the during the period of the experiment: Only 7.43\,\% of participants increased their public transport use frequency and decreased their car use frequency based on the categories shown in Figure \ref{fig:trip_freq}; contrary, around one third, did not report any changes and 20\,\% reported unchanged car use frequency but an increase in public transport use frequency. In total, around 20\,\% of our sample increased car use frequency during the period compared to the before period. Hence, we can conclude that  public transport use on a weekly basis was rather complementary to car use and not a substitute. Interestingly, Figure \ref{fig:trip_freq} presents evidence that the treatment of the \NineEuroTicket{} may have led to a slight increase in the sample's overall public transport usage: the share of regular public transport users in our sample grew by around 4 percentage points. However, this increase was in the trip frequency categories for one to three days per week.

\begin{table}[]
    \centering
    \begin{tabularx}{\textwidth}{X|rrr}
    \toprule
    Changes in car use frequency & \multicolumn{3}{c}{Changes in public transport use frequency}\\
    & Less & Unchanged & More \\
    \midrule
    Less &      3.88\,\%  &     8.88\,\%  &     7.43\,\% \\
    Unchanged &      8.78\,\% &     31.84\,\%   &   19.65\,\% \\
    More &     3.90\,\% &      9.23\,\%    &   6.36\,\% \\ 
\bottomrule
\end{tabularx}
    \caption{Cross tabulation of changes in car and public transport use frequency per week between the before and during the period.}
    \label{tab:trip_freq_changes}
\end{table}

In the second (during period) and third (after period) questionnaire, respondents stated whether they currently use public transport and the car more or less frequent compared to previous periods. Given our study's design, this leads to three cases:

\begin{itemize}
    \item comparing the during period to the before period
    \item comparing the after period to the during period
    \item comparing the after period to the before period
\end{itemize}

\begin{figure}
    \centering
    \includegraphics[width=\textwidth]{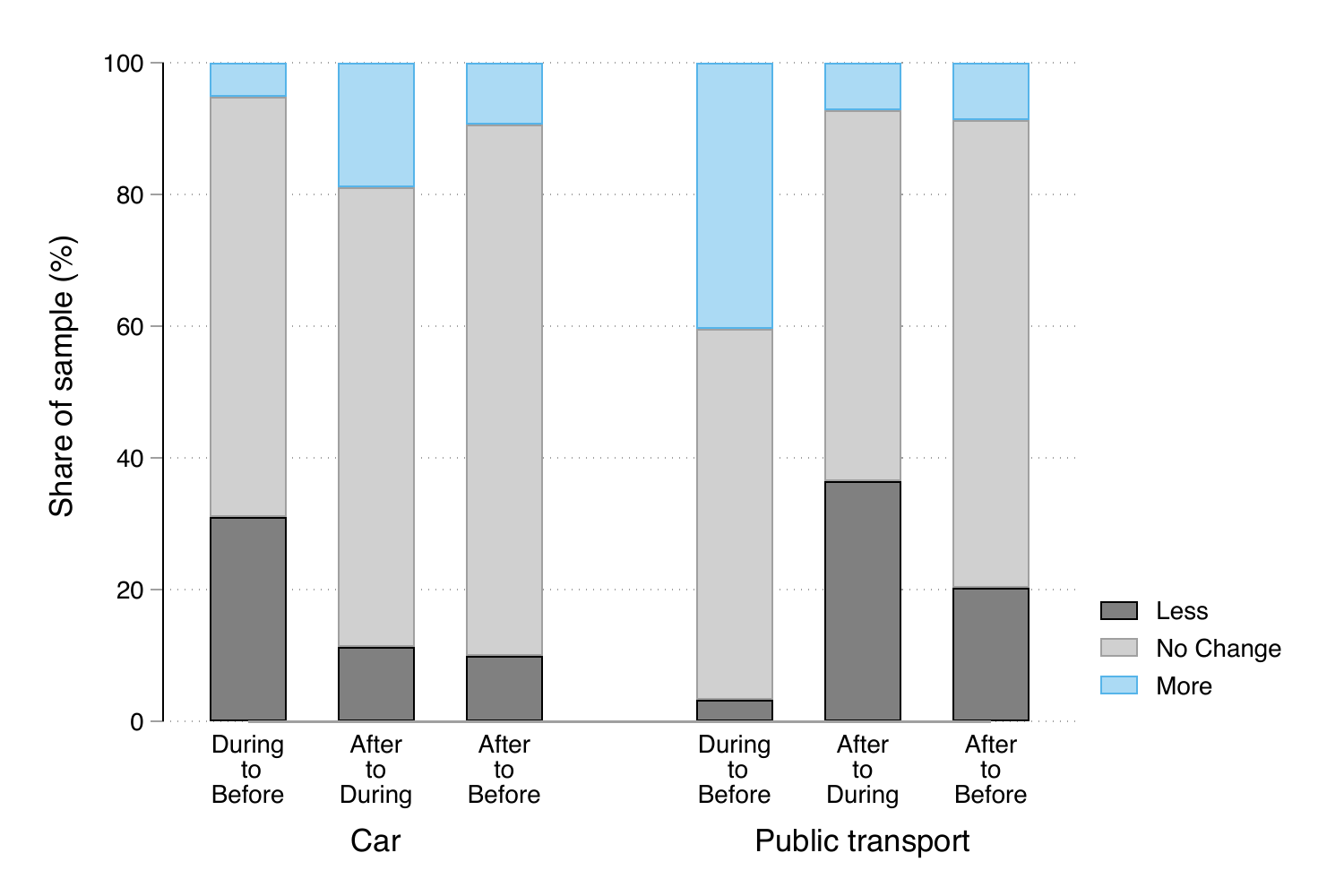}
    \caption{Stated changes in travel behavior between two periods of the experiment.}
    \label{fig:changes_tb}
\end{figure}

Figure \ref{fig:changes_tb} shows the responses. Regarding car use, the majority of participants did not report any changes throughout the study. A third stated that they used the car less in the during period compared to the before period, while around 20~\% stated that their car use increased in the after period compared to the during period. The presence of the \NineEuroTicket{}, which is partly intended to incentive changing from car to public transport, could be attributed to the stated changes. However, a likely \NineEuroTicket{} effect is more pronounced in public transport use: 40~\% reported an increase in public transport use in the during period compared to the before period, while slightly less than 40~\% reported less public transport use in the after period compared to the during period. The indication for activation of public transport caused by the \NineEuroTicket{} intervention as found in Figure \ref{fig:trip_freq} cannot be easily revealed from the qualitative responses shown in Figure \ref{fig:changes_tb}. However, the decrease in public transport use in the after period compared to the during period is five percentage points smaller than the increase in public transport use in the during period compared to the before period could be seen as such. 

\begin{figure}
    \centering
    \includegraphics[width=\textwidth]{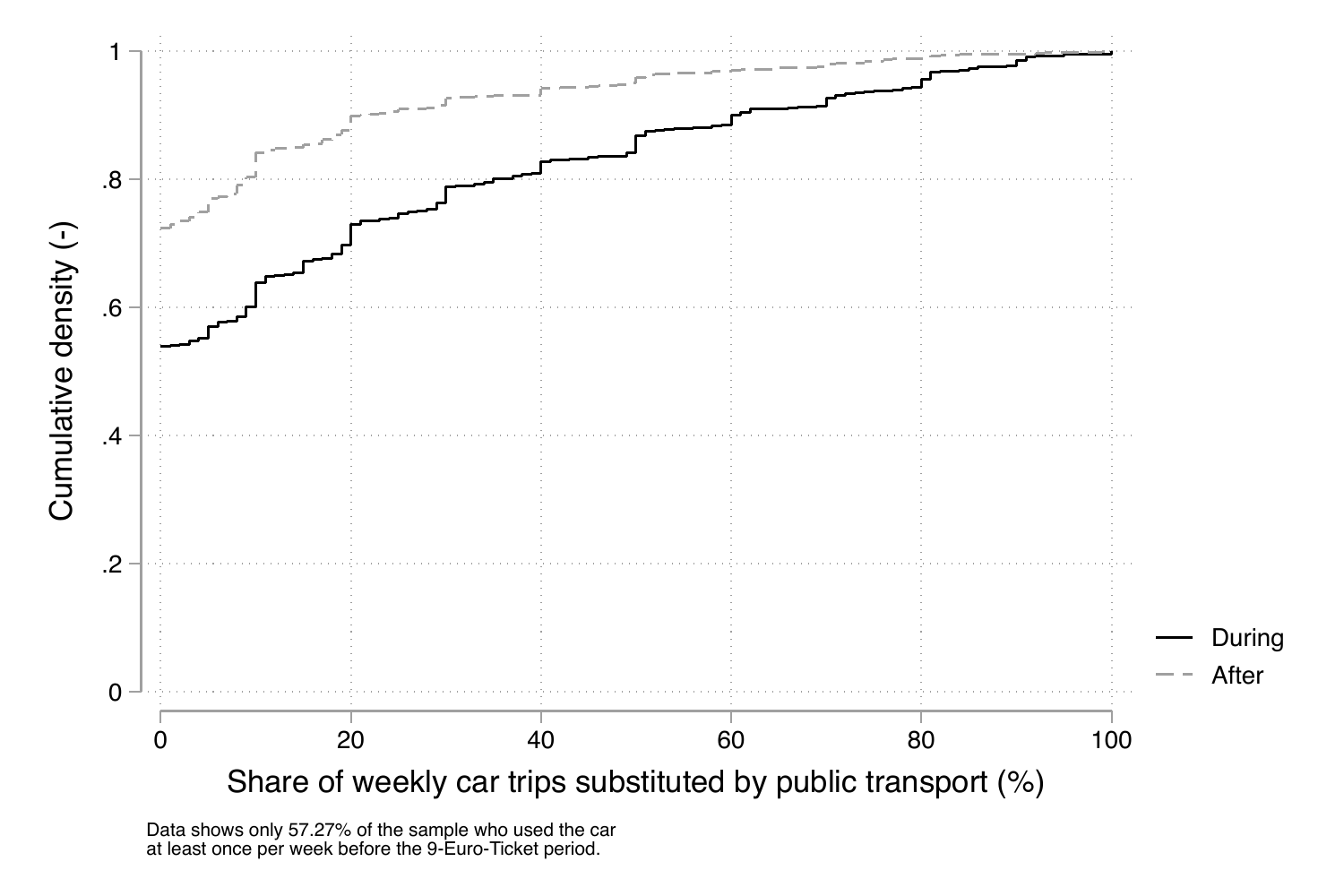}
    \caption{Cumulative density of the share of substituted car trips by public transport for all respondents who used the car at least once per week in May 2022.}
    \label{fig:share_of_trips}
\end{figure}

To better understand how study participants substitute car use by public transport, we asked respondents to provide a self-assessment of the share of substituted car trips per week. Substitution is only meaningful if the respondent used the car before the introduction of the \NineEuroTicket{} in May. Thus, Figure \ref{fig:share_of_trips} shows the self-reported substitution of car trips during and after the \NineEuroTicket{} period for all regular car users in May 2022. It can be seen that more than fifty percent of regular car users did not substitute any car trips, while only 20\,\% percent substituted more than 30\,\% of their weekly car trips during the \NineEuroTicket{} period. Nevertheless, after the \NineEuroTicket{} period, some respondents reported still substituting some car trips with around 10\,\% of the sample substituting more than 20\,\% of weekly car trips.

Overall, from the stated changes in travel behavior between the three study periods we can conclude that public transport use increased during the \NineEuroTicket{} period, but it did not substitute car use completely, suggesting induced demand.  In other words, the survey data suggests that respondents behaved more multimodal during the \NineEuroTicket{} experiment. Nevertheless, the data further present indications that the \NineEuroTicket{} activated some participants to use public transport more and continue to substitute car trips in the after period, i.e., after the treatment of the \NineEuroTicket{}. However, these findings from the survey will be corroborated and extended using a smartphone-based travel diary in the next section.

%% file: 05_app.tex
This section analyzes the data recorded by means of the \mbox{smartphone-based} travel diary as introduced in \mbox{Section \ref{sec:travel_diary}}. As described before, not all participants have concluded both the survey and tracking experiment. In this section, the considered \mbox{sub-sample} consists of 910 participants who recorded at least one trip on the smartphone application between \mbox{May 24, 2022,} and \mbox{September 30, 2022}. In other words, we investigate the travel behavior of all participants in this section irrespective of completing all questionnaires. Note that 920 participants successfully activated the app on their smartphones, but ten participants did not report any trips. Therefore, \mbox{socio-demographic} attributes relevant to travel behavior like age, income, gender, and place of residence for this \mbox{sub-sample} are not identical to the study sample reported in Section \ref{sec:final_sample}. However, we find no substantial differences to the characteristics reported in \mbox{Table \ref{tab:sample_info}} and \mbox{Figure \ref{fig:age_distribution}} and thus omit their presentation for the sake of conciseness. In summary, male and female participants are also represented almost equally by the tracking sample. Participants between the ages of 20 and 40 are slightly over-represented and most participants live in households with comparatively high incomes. Most participants live in the southeast of Germany, with a clear concentration in the Munich metropolitan area and thus within a densely connected public transport service area. In the subsequent analyses, we do only consider trips conducted within the geographic borders of Germany.

\begin{figure}
    \centering
    \includegraphics[width=\textwidth/2]{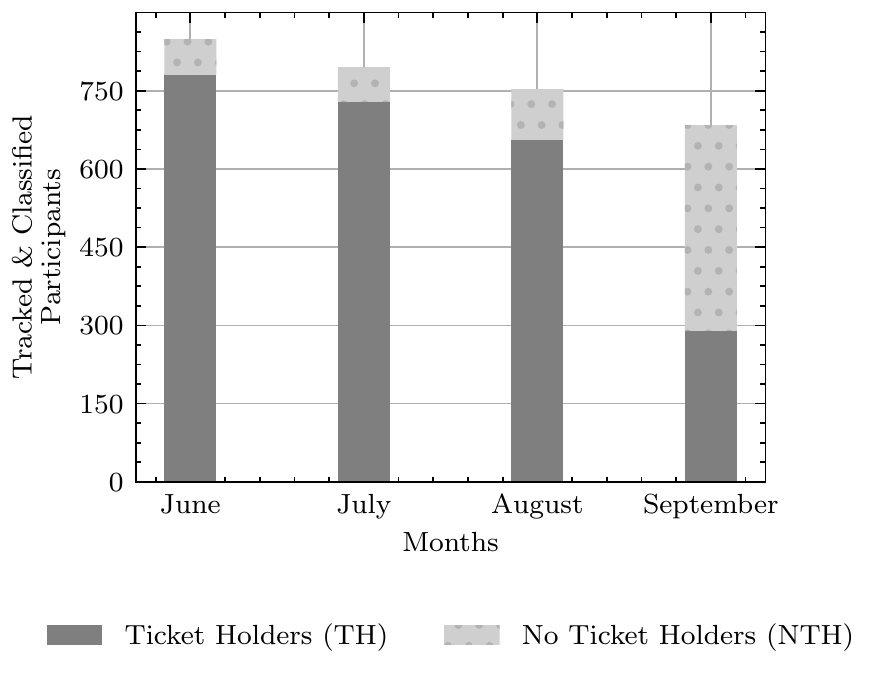}
    \caption{Travel pass ownership per month in the active tracking sample.}
    \label{fig:tracked_classified _participants}
\end{figure}

\mbox{Figure \ref{fig:tracked_classified _participants}} shows the travel pass ownership for all monthly active app users of our study. In the months of June, July, and August this refers to \NineEuroTicket{} ownership, while in September it refers to regular travel pass ownership. Holders of a public transport travel pass are denoted as ticket holders (TH) and \mbox{non-ticket holders} (NTH). It is important to note that \mbox{non-ticket holders} may also use public transportation on a \mbox{pay-per-trip} basis. It is apparent that during the \NineEuroTicket{}\ period, most of our active tracking participants did have a permanent subscription, whereas, after the end of the ticket subsidies, less than half of the active users owned a permanent ticket.

\begin{figure}
    \centering
    \includegraphics[width=\textwidth]{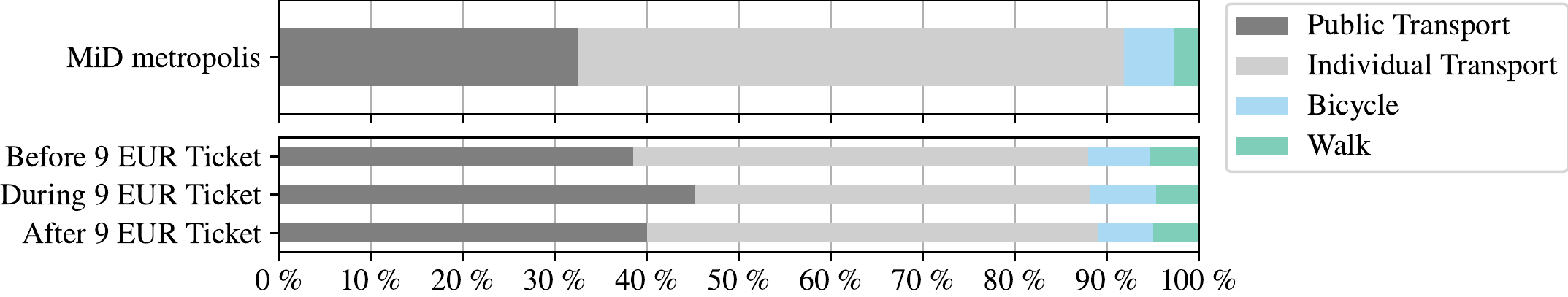}
    \caption{Modal share based on the traveled distance. From top to bottom: data derived from \gls{mid} (\cite{BundesministeriumfurVerkehrunddigitaleInfrastruktur.2018}) for metropolis regions; recorded data with our smartphone-based travel diary before, during, and after the \NineEuroTicket{}\ period}
    \label{fig:modal_share}
\end{figure}

To assess the impact of the \NineEuroTicket{}\ on our tracking sample's travel behavior we partition the collected data into three parts following the study design shown in Figure \ref{fig:study_design}: the first part includes all trips recorded before the \NineEuroTicket{}\ was introduced on June\ 1,\ 2022; the second part contains all recorded trips during the validity period of the \NineEuroTicket{}\ between June\ 1,\ 2022 and August\ 31,\ 2022; the third part includes all trips between September\ 1 and September\ 30,\ 2022 and represents the time period directly after the \NineEuroTicket{}. Note that only trips from within Germany are considered. \mbox{Figure \ref{fig:modal_share}} shows the modal share by travel distance of our tracking sample in each three study periods. For reference, we present the modal share for all metropolitan regions taken from the \gls{mid} survey. Reference data for Munich is only available for the city and the local transit district, but not the entire metropolitan area: Public transport share in the city is 36\,\% and in the transit district 28\,\%. To allow for a more concise presentation, we aggregate the means of transport as implemented in the smartphone-based travel diary in the following manner:

\begin{itemize}
    \item \emph{Public Transport}: subway, light rail, regional train, train, bus, tram
    \item \emph{Bicycle}: bicycle, bike-sharing, e-bicycle, kickscooter
    \item \emph{Individual Transport}: car, e-car, motorbike, taxi, uber, carsharing
    \item \emph{Walk}: walk
\end{itemize}

%\klaus{Figure 10: für München gibt es doch sicher aus der 2017 MID einen eigenen Modal split und nicht nur Metropolis}

Comparing the 2017 modal share from the \gls{mid} study to our collected data, we observe a higher use of public transport and a lower car use within all periods of the experiment. 
Comparing the modal shares of the three study periods, it can be seen that the introduction of the \NineEuroTicket{}\ had no clear impact on the share of distance covered by bicycles, but after the \NineEuroTicket{}\ ended, the share decreased slightly, most likely due to decreasing temperatures \citep{Heinen_Bike_Temperature} (the mean temperature in Munich was \SI{20.2}{\celsius} in the time range before and during the \NineEuroTicket{}\ and decreased to \SI{14.2}{\celsius} afterwards)\footnote{Calculated based on the public available LMU weather data for Munich \url{https://www.meteo.physik.uni-muenchen.de/DokuWiki/doku.php?id=wetter:stadt:messung}}. 
In contrast, the recorded share of walking clearly decreased in comparison between before and during the \NineEuroTicket{}\, albeit almost constant weather conditions. This could indicate that the hurdle to use public transport for smaller distances diminished for our sample group with the cheap ticket.

The observed shares of individual transport and public transport usage behave the other way around: before the \NineEuroTicket{}\ was introduced, the share of individual transport accounted for almost half of all traveled distances; it  decreased during the time of the \NineEuroTicket{} and increased again after the \NineEuroTicket{}\ period to a value slightly less than before the \NineEuroTicket{}. The share of public transport usage within our sample group increased with the \NineEuroTicket{}\ and it decreased when the \NineEuroTicket{}\ program ended but not as much as it increased before (see Figure \ref{fig:tracked_classified _participants}).

\begin{figure}
    \centering
    \includegraphics[width=\textwidth]{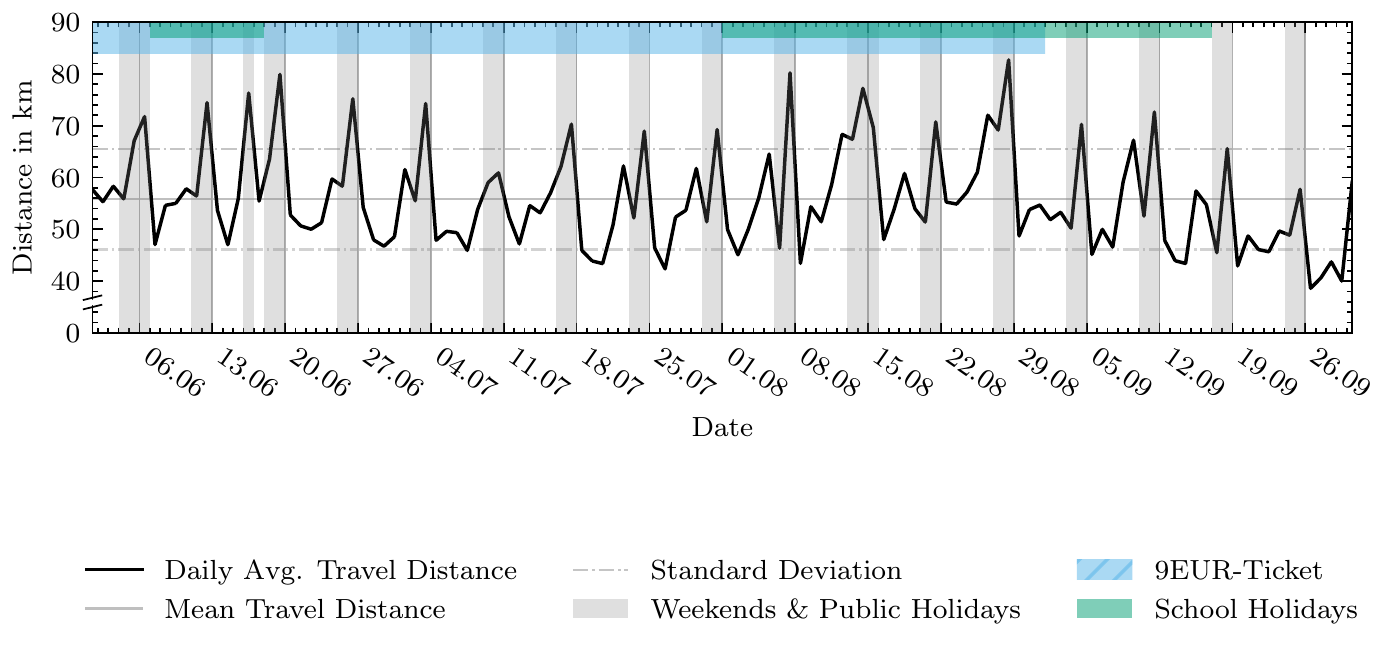}
    \caption{Average traveled distance per day and active user}
    \label{fig:daily_distance_avg}
\end{figure}

An overview of the daily traveled distances of participants that had at least one trip overall is given in \mbox{Figure \ref{fig:daily_distance_avg}}. Public holidays, weekends, and school holidays are indicated, revealing that a large part of the observation period of this study is essentially \mbox{non-working} days for a large share of the population. Potential impacts of the \NineEuroTicket{}\ on travel behavior are thus overlapped by the influences of holiday trips and other likely \mbox{infrequent} leisure activities. A reoccurring pattern is given by the increased daily travel distance on \mbox{non-working days}: in each period of the experiment, traveled distances peak on weekends. The highest daily travel distances are observed on weekends which happen to be school holidays during the \NineEuroTicket{}\ period. This indicates that leisure activities likely play a major role in the use of the \NineEuroTicket{}; a hypothesis that can be further substantiated when considering the drop in relative peak heights at the beginning of September after the end of the \NineEuroTicket{}. The last week of the observation period is the only week analyzed in which the \NineEuroTicket{}\ was not valid while there were simultaneously no school holidays. It is the only week providing data to compare a \mbox{non-holiday} situation with the \NineEuroTicket{}\ to a \mbox{non-holiday} period without the \NineEuroTicket{}. In this week, however, a significant drop in the average daily traveled distance is recognizable, substantiating the hypothesis that the \NineEuroTicket{}\ had a generally positive effect on traveled distances. This assumption is fortified by the fact that all but one weekend peaks during holidays and/or the \NineEuroTicket{} exceed the standard deviation of the time series, whilst they stay below it in the recorded week after the holidays. This effect is visible, whereas we can not give a clear measure of the additional impact of Oktoberfest (the Munich beer festival) from September 17, 2022, to October 3, 2022, and the reduced perception of COVID-19 in society referring to transportation numbers.

%\allister{@FTM: I agree, but shall we mention or comment that we cannot rule out that in the latter period, COVID quarantine and the Oktoberfest could also contribute to the observed travel behavior?}

\begin{figure}
    \centering
    \includegraphics[width=\textwidth]{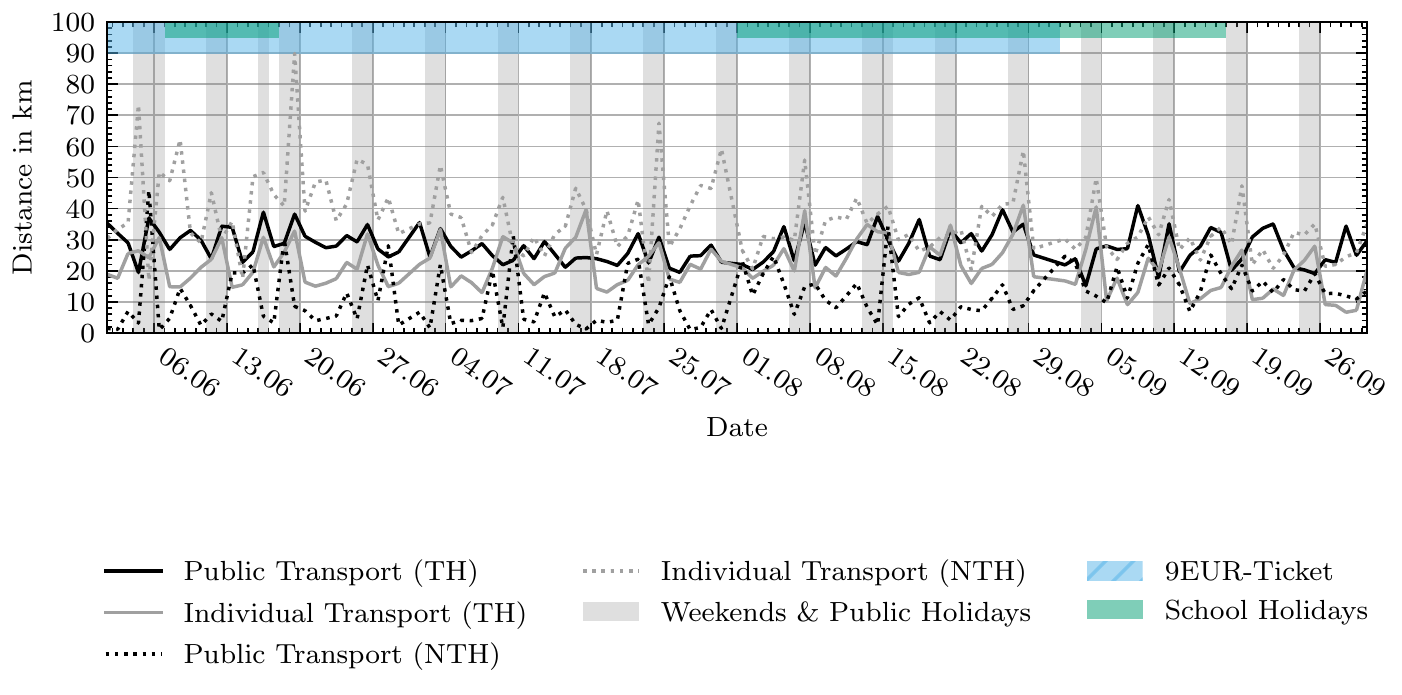}
    \caption{Average traveled distance per day and active user by means of transport and PT ticket ownership ((N)TH: (non) ticket holder}
    \label{fig:daily_distance_avg_mode_th}
\end{figure}

\mbox{Figure \ref{fig:daily_distance_avg_mode_th}} shows how the average daily traveled distance for individual and public transport varies for ticket holders and \mbox{non-ticket} holders as defined in Figure \ref{fig:tracked_classified _participants}. \mbox{Non-ticket} holders show a different travel behavior during the \NineEuroTicket{} phase and after this phase. During the ticket phase, in which the usage of public transport is almost free, they hardly travel using public transport. Instead, most travel distance is done by means of individual transport. The only exception to this observation is the sporadic usage of public transportation on weekends and public holidays which even then is significantly smaller than the peaks of individual transport of the same group. Only after the end of the \NineEuroTicket{}\ in September, \mbox{non-ticket} holders start to utilize public transport more often. While this seems like a noteworthy adverse effect at first glance, it is to be noted, that the size and composition of the \mbox{non-ticket} holder group can be assumed to change drastically after the end of the \NineEuroTicket{}\ (compare \mbox{Figure \ref{fig:tracked_classified _participants}}). While between June and August, only a small share of the participants chose not to buy the cheap \NineEuroTicket{}, less than half of the tracked participants bought a public transport travel pass in September.

%a greater heterogeneity of general attitudes toward public transport in the group of \mbox{non-ticket} holders. Hence, from \mbox{September\ 1, 2022} the differences between the two groups diminish.

\begin{figure}[htp]
\sbox\twosubbox{%
  \resizebox{\dimexpr.999\textwidth-1em}{!}{%
    \includegraphics[height=3cm]{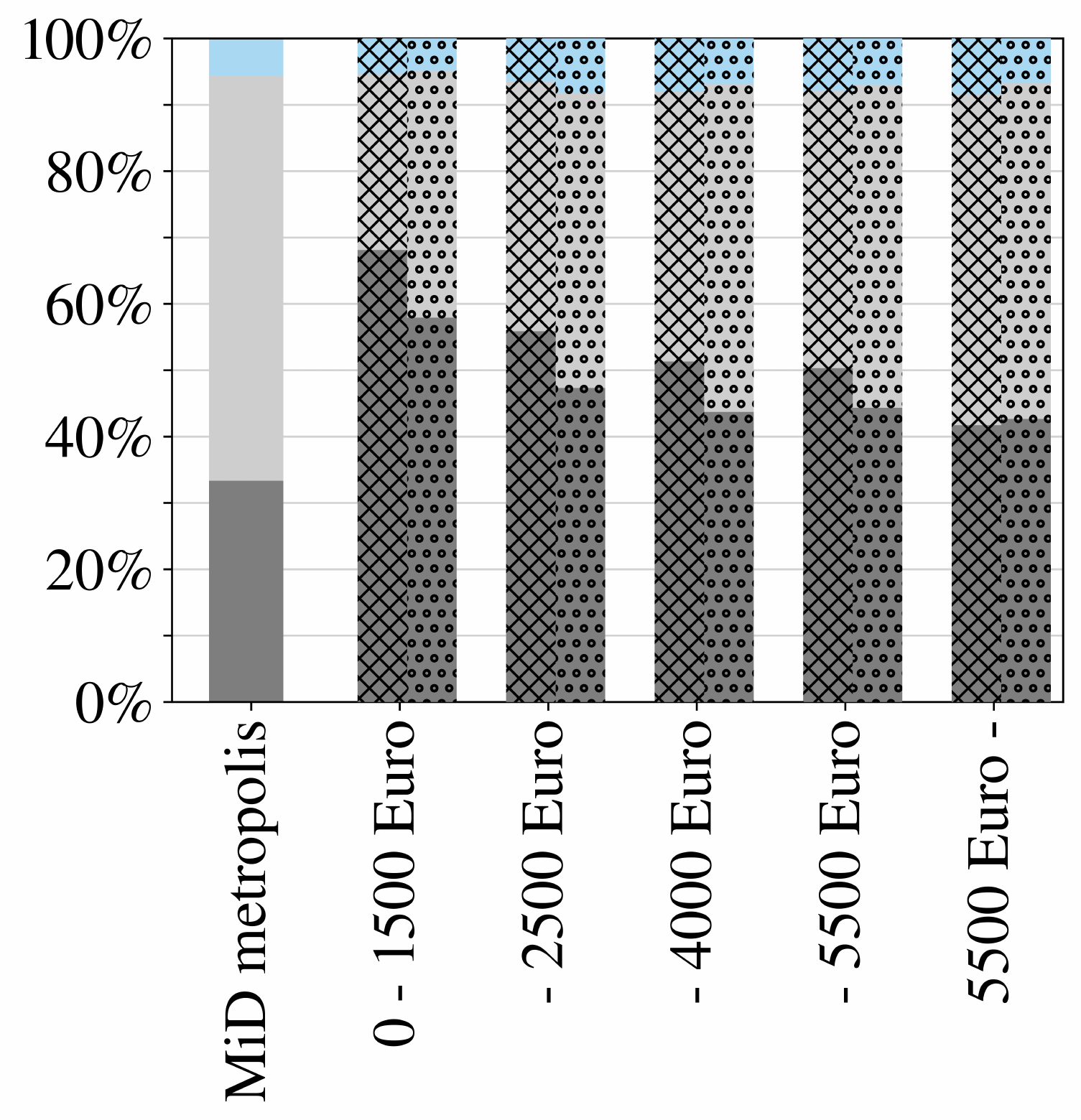}%
    \includegraphics[height=3cm]{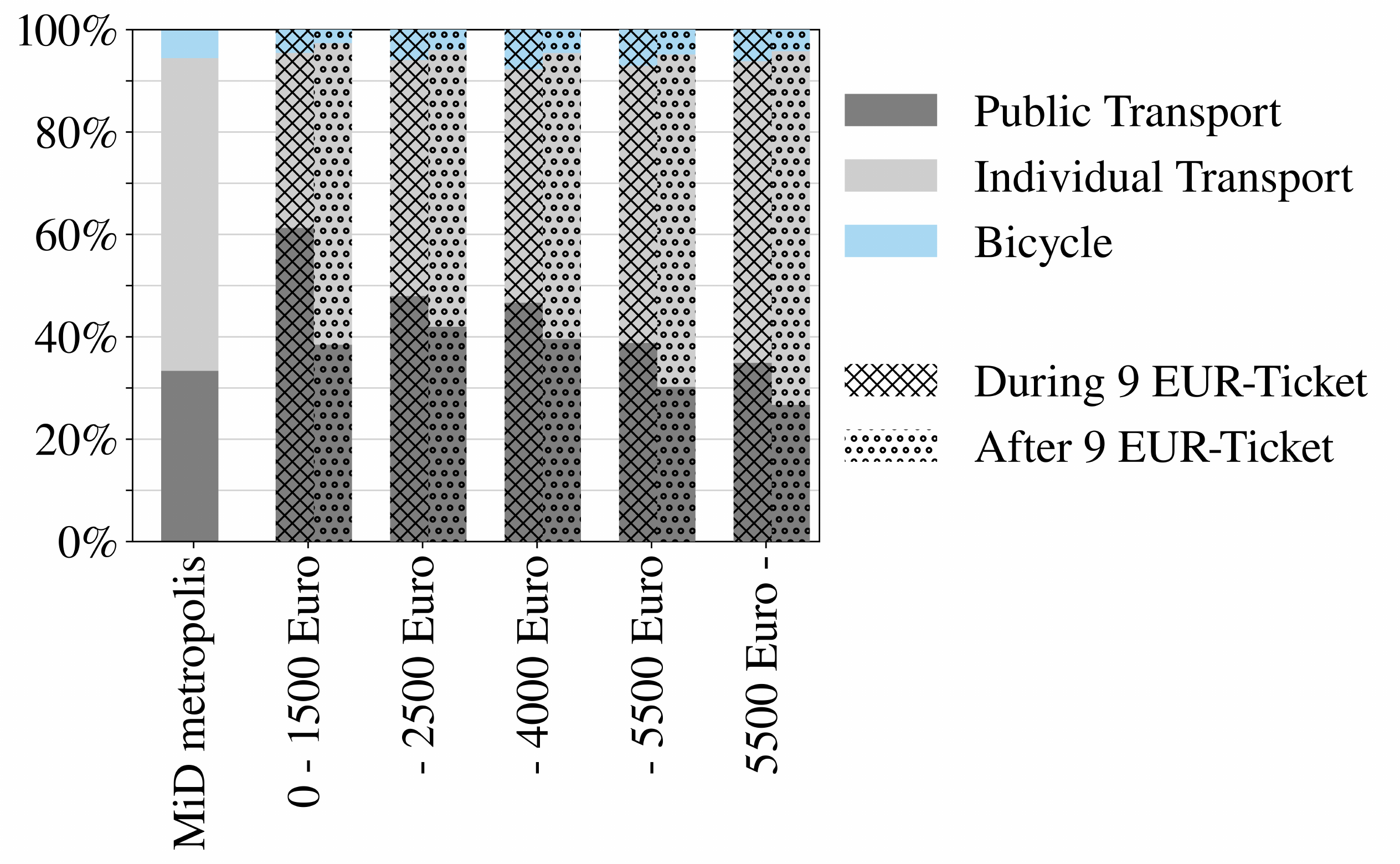}%
  }%
}
\setlength{\twosubht}{\ht\twosubbox}
\centering
\subcaptionbox{Working days\label{fig:MS_by_income_WD}}{%
  \includegraphics[height=\twosubht]{FIGURES/inc_wd_norm.pdf}%
}\quad
\subcaptionbox{\mbox{Non-working} days\label{fig:MS_by_income_NWD}}{%
  \includegraphics[height=\twosubht]{FIGURES/inc_we_norm.pdf}%
}
\caption{Modal share by income group}
\end{figure}

To assess the impact of the \NineEuroTicket{} on the travel behavior of different income groups within our tracking sample, we analyze the respective modal share in the during and after periods in Figures \ref{fig:MS_by_income_WD} and \ref{fig:MS_by_income_NWD}. In contrast to Figure \ref{fig:share_of_trips}, we do not analyze the period before the introduction of the \NineEuroTicket{}\ in this analysis. This is due to the further subdivision of trips by income groups that results in substantial differences in remaining total trip quantities, which would arguably lead to very small sample sizes. Instead, we use the \gls{mid} \cite{BundesministeriumfurVerkehrunddigitaleInfrastruktur.2018} as a reference and compare the results of our sample group to the national average for German metropolises. Due to the small contribution of walking to the total travel distance, this mode is not considered in this analysis. Values are normalized to the total distance covered by the remaining three means of transport. To account for the known differences in travel behavior between working and \mbox{non-working} days \cite{Gerike.2018}, we strictly distinguish between working days as well as weekends and public holidays (once more grouped as \mbox{non-working days}).

Generally, public transport use is higher on working days than on \mbox{non-working} days. The difference amounts to roughly ten percentage points throughout income groups. For both day types, public transport usage decreases with increasing income. Considering \mbox{non-working} days, the largest decrease in public transit usage of 25 percentage points occurs for the group 0-1500~Euro of monthly household income. All other groups' public transport usage on \mbox{non-working} days decreases by about 10 percentage points. On working days, all groups but the one with the highest household income show an equivalently declining public transport usage once the \NineEuroTicket{} was terminated. The group with a monthly income of more than \mbox{5,500 EUR} is the only one without the relevant impact of the ticket on working days. It is worth noting that the mode choice of the latter group on \mbox{non-working} days, in contrast to working days, exhibits a change between the periods, as public transit usage was higher when the \NineEuroTicket{} was available.

%\klaus{activities hat man nicht angeschaut??? man beschreibt sie aber bei der smartphone app als feature}

We conclude that the travel behavior revealed with the smartphone-based travel diary shows a similar trend as the stated responses in Section \ref{sec:sp_travel}. However, the tracking allows a more precise quantification of effects: the observed shift in the modal share by travel distance from car to public transport is around five percentage points. Results also suggest generally larger travel distances on weekends during the \NineEuroTicket{} period, particularly by ticket holders who use public transport. The presented data shows that the \NineEuroTicket{} has a stronger impact on increasing the use of public transport among lower-income groups, as compared to the highest-income group, who barely change their travel behavior on working days.

%Participants from the highest income group barely change their travel behavior on working days. The other income groups render an increasing usage of public transport during the \NineEuroTicket{} phase for both day types. Data suggests a stronger effect of the \NineEuroTicket{} on public transport usage for lower-income groups than for higher.

%All income groups both on working and non-working days, except the highest income group on workdays, show a reduction in public transport usage after the \NineEuroTicket{} period suggesting that participants from the highest income group barely changed travel (i.e., commuting) behavior during the week.

%% file: 06_consumer.tex
The \NineEuroTicket{} is an unprecedented intervention to the German transportation system. As there has never been something similar before, it is of interest to understand important dimensions of the \NineEuroTicket{} as a consumer product too. Thus, we study in this section the public opinions of the \NineEuroTicket{} (Section \ref{sec:public_opinion}), its support (Section \ref{sec:support}), buying intention (Section \ref{sec:intention}), as well as the willingness-to-pay (henceforth WTP) for a successor ticket (Section \ref{sec:wtp}). 

\subsection{Public opinion}\label{sec:public_opinion}

\begin{figure}
    \centering
    \includegraphics[width=\textwidth]{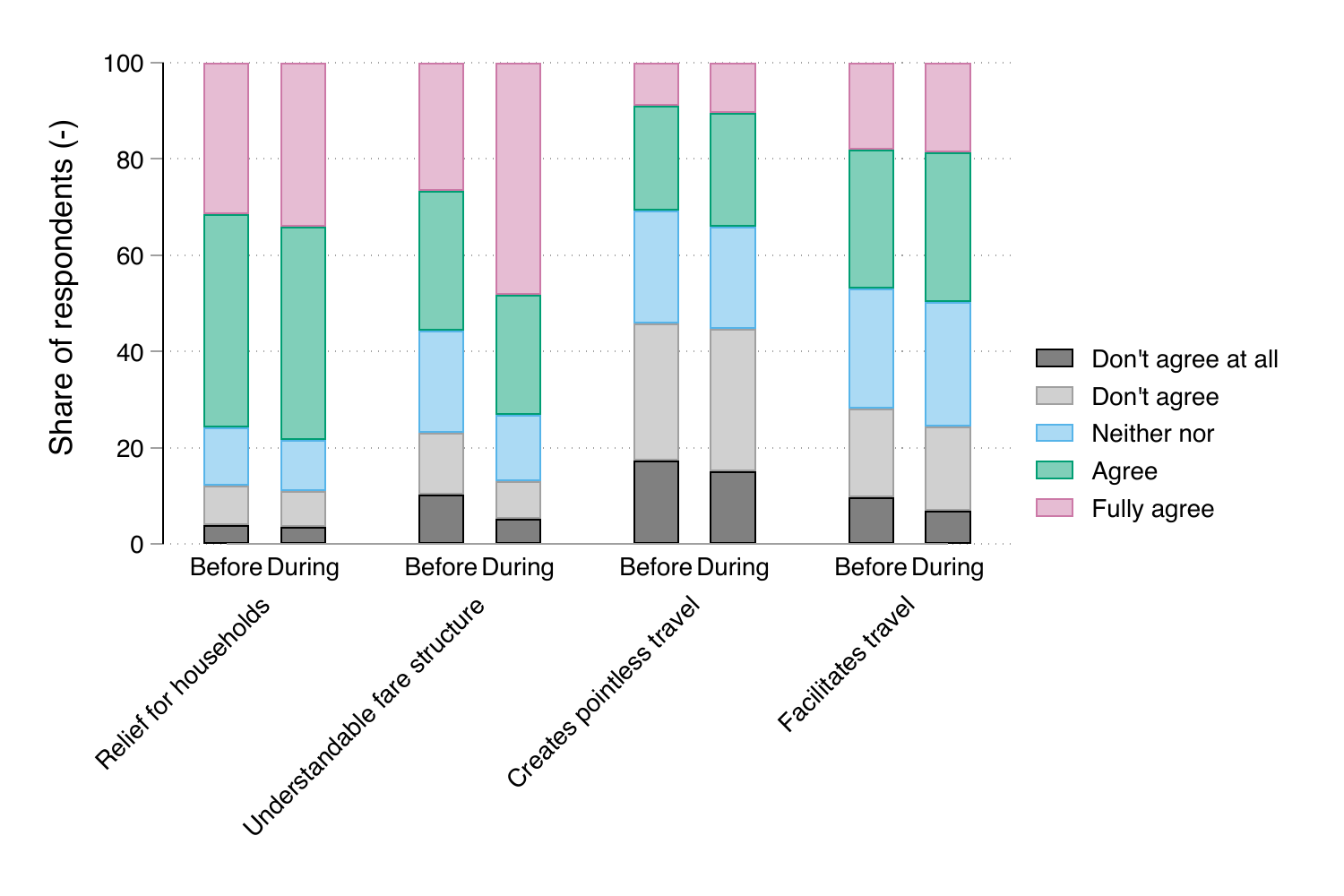}
    \caption{Participants opinions about the \NineEuroTicket{} in wave one and two}
    \label{fig:statements_w1w2}
\end{figure}

In the first (i.e., before the introduction) and second (i.e., during the \NineEuroTicket{}) questionnaire of the panel survey, respondents stated their opinion towards the \NineEuroTicket{} using several statements on a disagree-agree scale. \mbox{Figure \ref{fig:statements_w1w2}} shows selected statements that have been frequently discussed in the public. We find that respondents have positive attitudes towards the \NineEuroTicket{} and that this positive view is persistent over time. For instance, we find that the majority of participants perceives the \NineEuroTicket{} as a relief for households as it had been intended by the federal government. Interestingly, we find an increasing agreement with the statement of the \NineEuroTicket{} is increasing the comprehensibility of the fare structure of public transport in Germany from the before to the during period. Here, participants seem to become more aware of this advantage during the availability of the ticket. Only one third of participants agrees with the statement that the ticket leads to pointless travel and this does also not change substantially once the ticket is introduced. Half of participants agrees that the \NineEuroTicket{} facilitates traveling in Germany by public transport. 

Furthermore, in the second questionnaire in the during period participants are asked to indicate what they consider to be the greatest benefit of the \NineEuroTicket{} (they could choose between price, flexibility and comprehensibility or indicate another advantage). Here, 54.3\,\% mentioned price, while 24.5\,\% mentioned flexibility and 16.3\,\% mentioned simplicity. 

\subsection{Support}\label{sec:support}

\begin{figure}
    \centering
    \includegraphics[width=12cm]{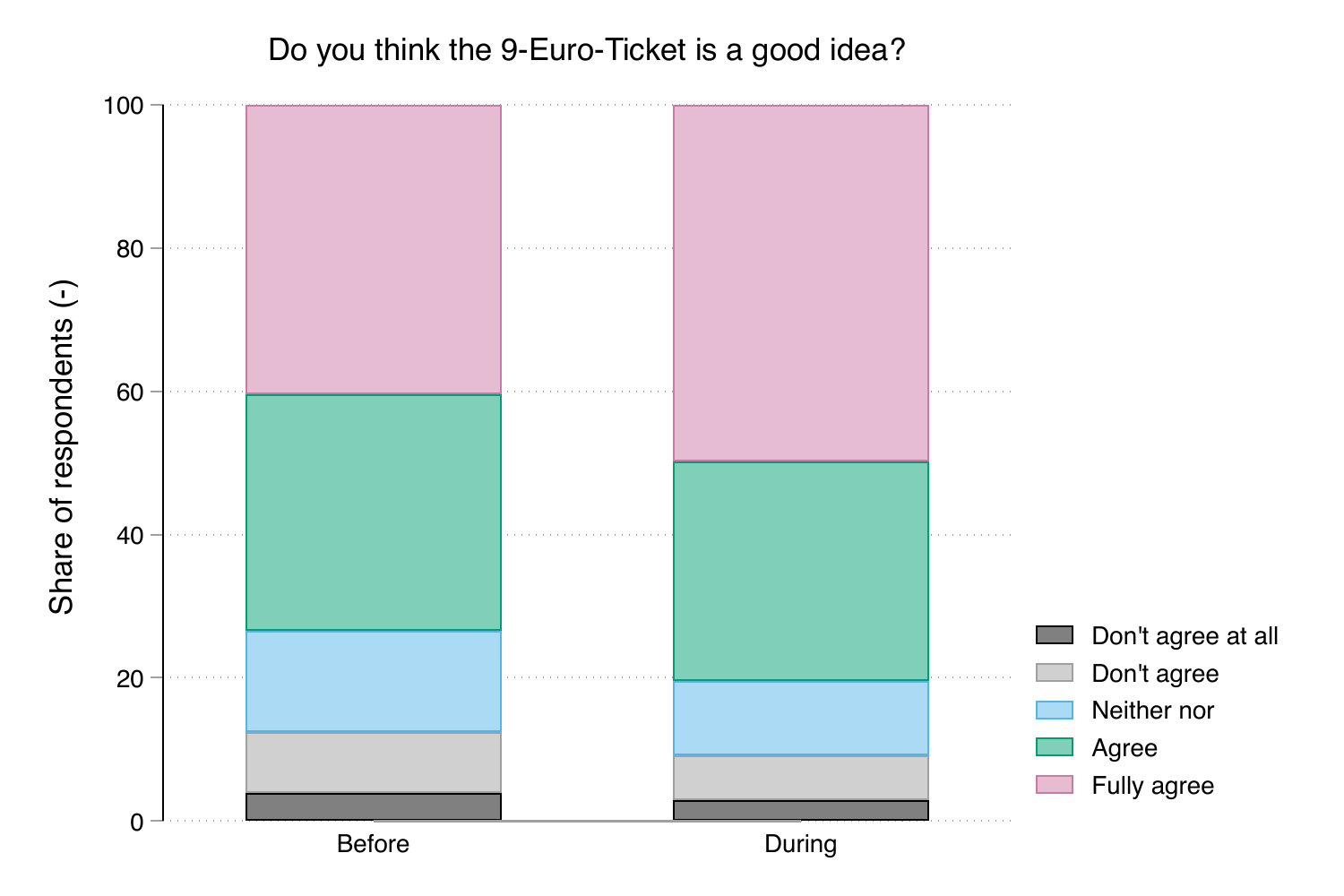}
    \caption{Support for the ticket in the before and during period of the experiment.}
    \label{fig:support}
\end{figure}

Based on the positive perception of the \NineEuroTicket{} in the first (before) and second (during) questionnaire we can expect high levels of support for the ticket. In our study, we assess support for the \NineEuroTicket{} using a five-point agree-disagree scale for the statement whether they consider the \NineEuroTicket{} is a good idea. Looking at the results in \mbox{Figure \ref{fig:support}}, we find that the majority of participants at least support the ticket before its introduction. This already high support for the ticket substantially increases in the during period. In our sample, we find that only 10 percent disagree with the statement that the ticket is a good idea.

Interestingly, we find that age is a significant determinant of support, where older participants support the ticket significantly less. In addition to this age effect, we find a slightly significant income effect in the before period: participants in the second highest and highest income category support the ticket significantly less compared the lowest income category. This income effect disappears in the second wave, where we find no difference in support between income levels anymore. 
% \citep{cantner_nation-wide_2022}
% \citep{loder_nation-wide_2022}

\subsection{Intention to buy it and actual buying behavior} \label{sec:intention}

Participants were asked before the introduction of the \NineEuroTicket{} about their preferences to obtain this. Table \ref{tab:intent_actual} compares the stated intention and the actual adoption in each of the three months of the \NineEuroTicket{} period.

It can be seen in Table \ref{tab:intent_actual} that from June to July and August, the share of already bought tickets is reduced by around 20 percentage points, while the share of participants having the intention to buy it increases by the same amount. As this question has been asked before the introduction of the \NineEuroTicket{}, we can conclude that a large part of our sample bought the \NineEuroTicket{} every month individually. We also find that in our sample the overall interest in the ticket seemed to have declined by a few percentage points from June from 73.22\,\% to 70.47\,\% in August. Overall we find that a large and consistent share (around 85\,\% in each month) of participants that indented to buy the ticket actually bought it. 

\begin{table}[]
\begin{tabularx}{\textwidth}{X|r|rr|r|rr|r|rr}
\toprule
& \multicolumn{6}{c}{}{Actual buying behavior (Shares of the sample)}\\ \cline{2-10}
 & \multicolumn{3}{c|}{June} & \multicolumn{3}{c|}{July} & \multicolumn{3}{c}{August} \\ \cline{2-10} 
 Buying intention category & \multicolumn{1}{c}{Total} & \multicolumn{1}{c}{No}  & \multicolumn{1}{c}{Yes}  & \multicolumn{1}{|c}{Total} & \multicolumn{1}{c}{No}  & \multicolumn{1}{c}{Yes}  & \multicolumn{1}{|c}{Total} & \multicolumn{1}{c}{No}  & \multicolumn{1}{c}{Yes}  \\ \midrule
Already bought      & 41.32 & 0.00 & 100.00  & 22.29 & 0.00 & 99.44  & 21.73 & 0 & 100.00 \\
Intention to buy    & 28.09 & 14.76 & 85.24 & 47.23 & 15.09 & 84.91  & 46.33 & 13.98 & 86.02 \\
No intention        &25.38 & 78.59 & 21.41  & 23.15 & 76.01 & 23.99 & 23.02 & 78.44 & 21.56 \\
Do not know         & 5.22 & 57.53 & 43.47  & 7.33 & 53.92 & 46.08  & 8.92 & 52.42 & 47.58 \\
\midrule
Total & 100.00 & 27.09 & 72.91 & 100.00 &  28.65 & 71.35 & 100.00 & 29.53 & 70.47 \\
\bottomrule
\end{tabularx}
\caption{Buying intention vs. actual (stated) buying behavior for the three months of the \NineEuroTicket{} period.}
\label{tab:intent_actual}
\end{table}

\subsection{Willingness-to-pay for a successor ticket} \label{sec:wtp}

We assess the willingness-to-pay (WTP) for a successor to the \NineEuroTicket{} in the during and after period of \NineEuroTicket{}. Given the substantially different income levels in the Munich-based and nation-wide panel as seen in Table \ref{tab:sample_info}, we report the results separately for each panel to accommodate that higher incomes correlate with higher WTP.

\begin{figure}
    \centering
    \includegraphics[width=\textwidth]{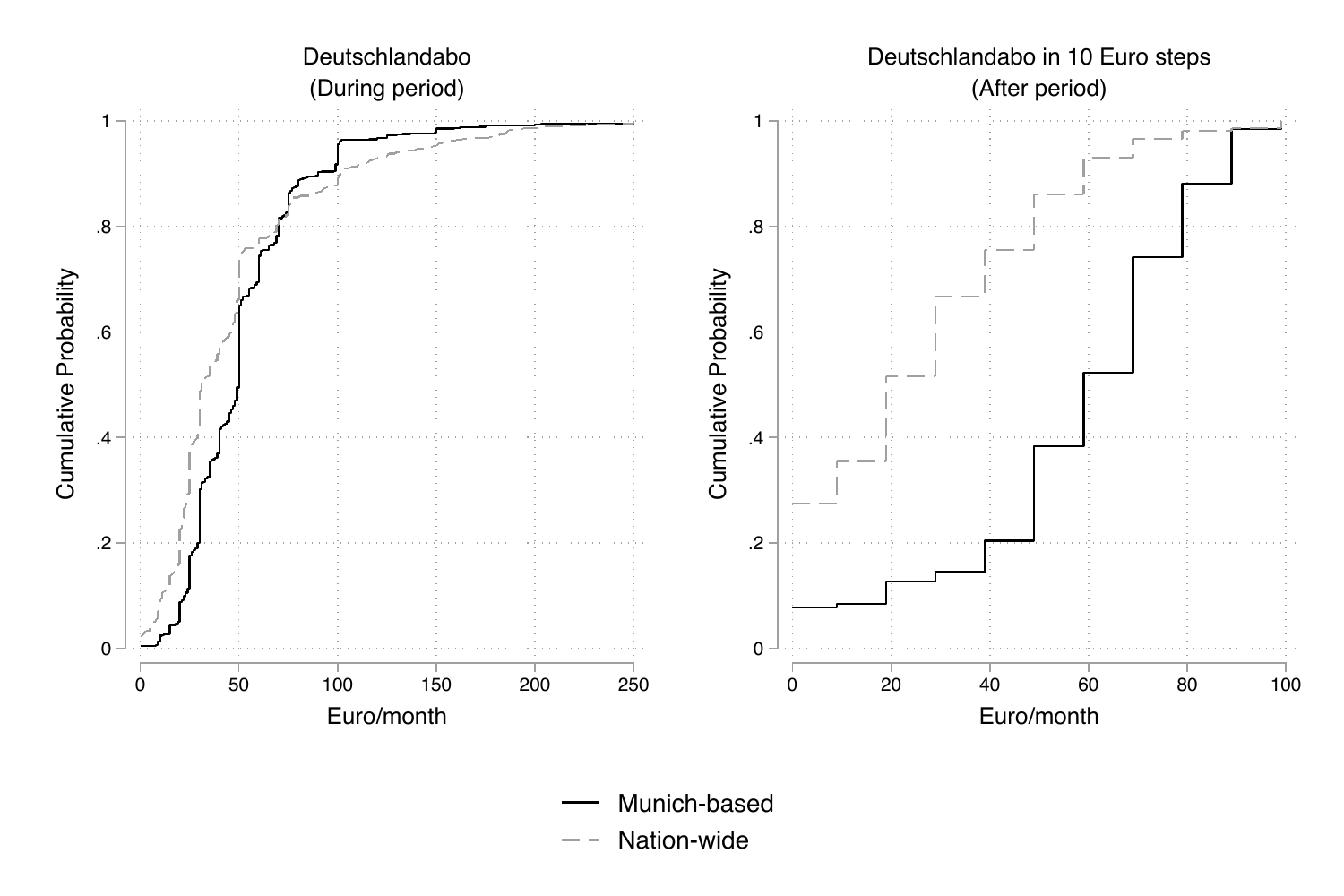}
    \caption{Cumulative distributions of the willingness-to-pay for a successor ticket to the \NineEuroTicket{}. On the left, Deutschlandabo on a continuous scale, and Deutschlandabo on a discrete scale on the right.}
    \label{fig:wtp}
\end{figure}

\begin{figure}
    \centering
    \includegraphics[width=\textwidth]{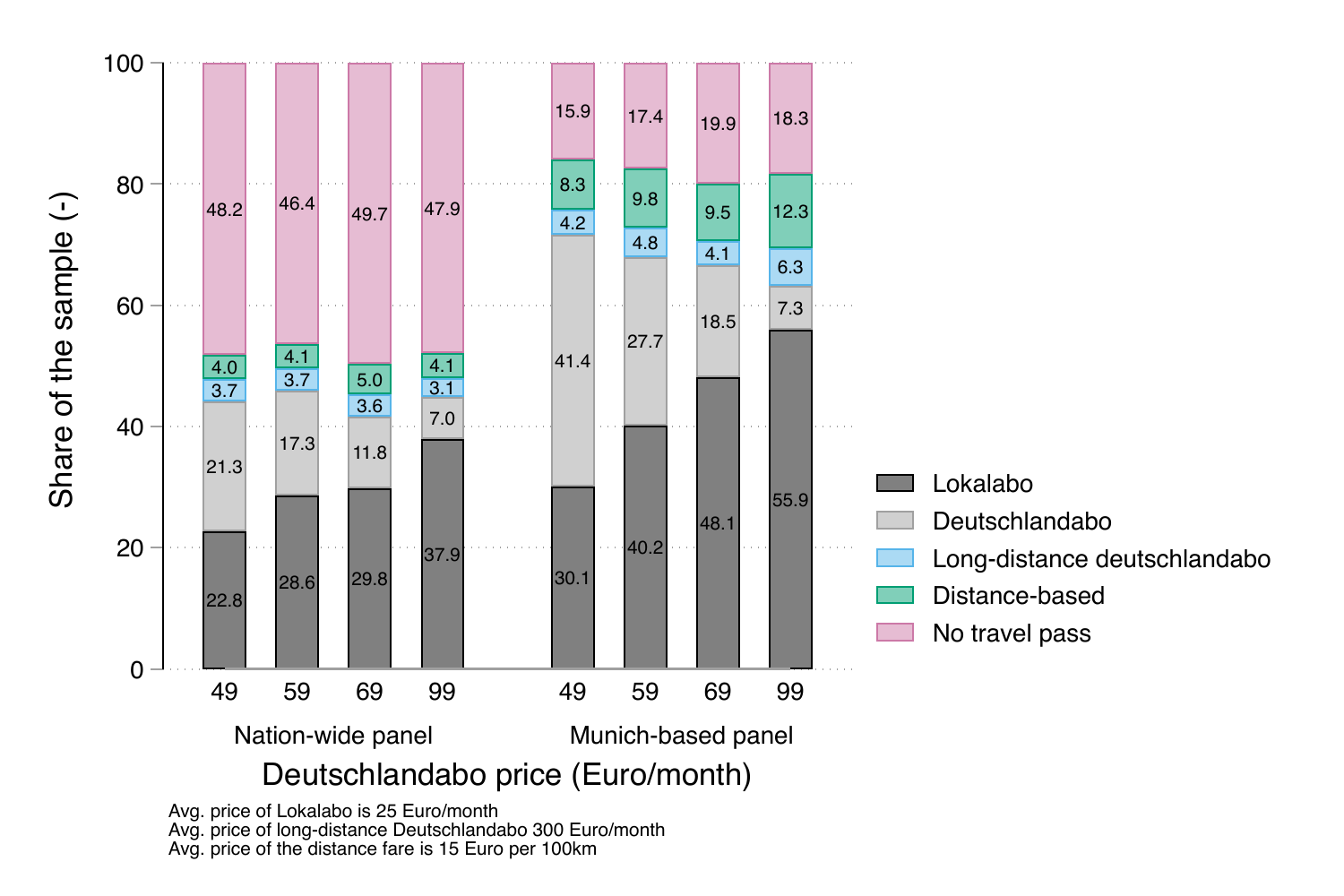}
    \caption{Shares of travel pass ownership choices in the discrete-choice experiment as a function of the price for the \NineEuroTicket{} successor ticket.}
    \label{fig:dce}
\end{figure}

In the second questionnaire in the during period, study participants were asked to state their maximum WTP for a nation-wide travel pass for all local public transport services, the so-called \textit{Deutschlandabo}. In the third questionnaire in the after period, respondents were asked in a twofold way to state their consumer preferences for a successor to the \NineEuroTicket{}, i.e., the \textit{Deutschlandabo}: first, based on the ongoing public debate on the pricing of the ticket, the maximum willingness-to-pay for a successor was asked on a discrete scale from 9 to 99\,Euro at an 10\,Euro interval; second, using the publicly discussed pricing points for the successor ticket a discrete-choice experiment with different ticket alternatives has been designed (see Section \ref{sec:survey} for further details). Figure \ref{fig:wtp} shows the distributions of the three direct responses in maximum WTP, while Figure \ref{fig:dce} shows the outcomes of the discrete-choice experiment for different price levels of the \textit{Deutschlandabo}.

Figure \ref{fig:wtp} shows that in both samples, only 10\,\% are willing to pay 100 Euro or more for a successor in the during period of the \NineEuroTicket{} with a median WTP of 30\,Euro/month for the nation-wide panel and 50\,Euro/month for the Munich-based panel. Interestingly, the median WTP increases for the Munich-based panel in the after period  to 59\,Euro/month, while being unchanged for the nation-wide sample. The average WTP for a Deutschlandabo as a successor of the \NineEuroTicket{} is 50.45~Euro which is close to the current average price paid for travel pass in Germany (around 55~Euro/month).

We compare the WTP for the Deutschlandabo based on the continuous scale (left side of Figure \ref{fig:wtp}) by socio-economic attributes using a linear regression model on all complete observations. Table \ref{tab:ols} shows the resulting estimates. We find that the average WTP is around 33~Euro which translates to roughly 1~Euro per day. We find no significant differences between males and females, Intuitively, we find that that having an income higher than 1,500\,Euro per month increases the WTP by 10~Euro to 15~Euro compared to the lowest income group (see Table \ref{tab:sample_info}). In our sample, we find no age effect. In addition, using the car frequently does not impact the WTP, but being a frequent public transport user before the introduction of the \NineEuroTicket{} increases the WTP by around 16~Euro. We further find that residing in an urban metropolis according to the RegioStar7 classification (see Table \ref{tab:sample_info}) decreases the WTP by around 5 Euro (almost statistically significant at the five percent level of statistical significance). This effect can be attributed to the fact that usually the price of a travel pass increases with distance to the core urban area. In other words, residents of this area are used to cheaper travel passes compared to residents in surrounding more suburban areas. Last, we control for the differences in the two panels and find that the WTP in the nation-wide panel is around 6 Euro less compared to the Munich-based panel, which can be explained by the higher income levels in and around Munich.

\def\sym#1{\ifmmode^{#1}\else\(^{#1}\)\fi}
\begin{table}[]
    \centering
\begin{tabularx}{\textwidth}{X|cc}
\toprule
Estimate & Model parameter & \textit{t}\\
\midrule
Constant      &       33.44\sym{***}&      (6.49)\\
Gender : Male (base)    &                    &        \\
Gender : Female    &       0.11         &      (0.03)\\
Gender : Diverse     &      -14.10\sym{**} &     (-2.66)\\
Net household income : $<$ 1,500 Euro / month (base) &                    &         \\
Net household income : 1,500 to 2,490 Euro / month &       11.22\sym{**} &      (3.06)\\
Net household income : 2,500 to 3,999 Euro / month&       13.54\sym{***}&      (3.77)\\
Net household income : $>$ 4000 Euro / month  &       15.72\sym{***}&      (3.66)\\
Age : 18 -- 24 (base)    &                    &         \\
Age : 25 -- 34    &       3.42         &      (0.90)\\
Age : 35 -- 44   &       1.11         &      (0.26)\\
Age : 45 -- 54    &       4.20         &      (0.93)\\
Age : 55 -- 64   &      -2.59         &     (-0.65)\\
Age : 65 -- 74    &      -1.13         &     (-0.25)\\
Age : $>$ 75    &       5.08         &      (0.47)\\
Uses the car more than one day per week   &       1.57         &      (0.43)\\
Uses public transport more than one day per week    &       16.27\sym{***}&      (5.39)\\
Living in urban metropolis (according to RegioStar7)     &      -5.60         &     (-1.91)\\
Participant is part of the nation-wide panel&      -6.40\sym{*}  &     (-2.48)\\
\midrule
\(N\)       &        1,333         &            \\
\(R^2\)       &        0.077         &            \\
\bottomrule
\multicolumn{3}{l}{\footnotesize \textit{t} statistics in parentheses}\\
\multicolumn{3}{l}{\footnotesize \sym{*} \(p<0.05\), \sym{**} \(p<0.01\), \sym{***} \(p<0.001\)}\\
\end{tabularx}
    \caption{Ordinary least squares estimates of the willingness-to-pay model}
    \label{tab:ols}
\end{table}
 
Figure \ref{fig:dce} shows the outcomes of the discrete-choice experiment conducted in the after period of the \NineEuroTicket{} experiment. It can be seen that with an increase in the price for the successor ticket, the participants' interest declines substantially. Interestingly, while in the nation-wide panel participants choose with increasing prices for the \NineEuroTicket{} successor the cheaper Lokalabo (local travel pass) instead, in the Munich-based panel  the shares of ``no travel pass'' and the ``distance-based pricing'' increase. The latter could be attributed to having the successor ticket as an ``accessoire'' if reasonably priced, but not as a mobility tool. Generally, when transit agencies offer a cheap local travel pass (Lokalabo), the travel pass ownership shares will substantially increase compared to existing travel pass ownership levels as shown in Table \ref{tab:sample_info}.

%
%
%

%% file: neunern_main.bbl
\begin{thebibliography}{63}
\expandafter\ifx\csname natexlab\endcsname\relax\def\natexlab#1{#1}\fi
\providecommand{\url}[1]{\texttt{#1}}
\providecommand{\href}[2]{#2}
\providecommand{\path}[1]{#1}
\providecommand{\DOIprefix}{doi:}
\providecommand{\ArXivprefix}{arXiv:}
\providecommand{\URLprefix}{URL: }
\providecommand{\Pubmedprefix}{pmid:}
\providecommand{\doi}[1]{\href{http://dx.doi.org/#1}{\path{#1}}}
\providecommand{\Pubmed}[1]{\href{pmid:#1}{\path{#1}}}
\providecommand{\bibinfo}[2]{#2}
\ifx\xfnm\relax \def\xfnm[#1]{\unskip,\space#1}\fi
%Type = Article
\bibitem[{Adler and van Ommeren(2016)}]{adler_does_2016}
\bibinfo{author}{Adler, M.W.}, \bibinfo{author}{van Ommeren, J.N.},
  \bibinfo{year}{2016}.
\newblock \bibinfo{title}{Does public transit reduce car travel externalities?
  {Quasi}-natural experiments' evidence from transit strikes}.
\newblock \bibinfo{journal}{Journal of Urban Economics} \bibinfo{volume}{92},
  \bibinfo{pages}{106--119}.
%Type = Techreport
\bibitem[{AG and VDV(2022)}]{deutsche_bahn_ag_abschlussbericht_2022}
\bibinfo{author}{AG, D.B.}, \bibinfo{author}{VDV}, \bibinfo{year}{2022}.
\newblock \bibinfo{title}{Abschlussbericht zur bundesweiten {Marktforschung}}.
\newblock \bibinfo{type}{Technical Report}.
%Type = Article
\bibitem[{Alessandretti et~al.(2018)Alessandretti, Sapiezynski, Sekara, Lehmann
  and Baronchelli}]{alessandretti_evidence_2018}
\bibinfo{author}{Alessandretti, L.}, \bibinfo{author}{Sapiezynski, P.},
  \bibinfo{author}{Sekara, V.}, \bibinfo{author}{Lehmann, S.},
  \bibinfo{author}{Baronchelli, A.}, \bibinfo{year}{2018}.
\newblock \bibinfo{title}{Evidence for a conserved quantity in human mobility}.
\newblock \bibinfo{journal}{Nature Human Behaviour} \bibinfo{volume}{2},
  \bibinfo{pages}{485--491}.
\newblock \DOIprefix\doi{10.1038/s41562-018-0364-x}.
%Type = Article
\bibitem[{Allström et~al.(2017)Allström, Kristoffersson and
  Susilo}]{allstrom_smartphone_2017}
\bibinfo{author}{Allström, A.}, \bibinfo{author}{Kristoffersson, I.},
  \bibinfo{author}{Susilo, Y.}, \bibinfo{year}{2017}.
\newblock \bibinfo{title}{Smartphone based travel diary collection: experiences
  from a field trial in {Stockholm}}.
\newblock \bibinfo{journal}{Transportation Research Procedia}
  \bibinfo{volume}{26}, \bibinfo{pages}{32--38}.
\newblock \URLprefix
  \url{https://linkinghub.elsevier.com/retrieve/pii/S2352146517308657},
  \DOIprefix\doi{10.1016/j.trpro.2017.07.006}.
%Type = Article
\bibitem[{Anderson(2014)}]{anderson_subways_2014}
\bibinfo{author}{Anderson, M.L.}, \bibinfo{year}{2014}.
\newblock \bibinfo{title}{Subways, {Strikes}, and {Slowdowns}: {The} {Impacts}
  of {Public} {Transit} on {Traffic} {Congestion}}.
\newblock \bibinfo{journal}{American Economic Review} \bibinfo{volume}{104},
  \bibinfo{pages}{2763--2796}.
\newblock \DOIprefix\doi{10.1257/aer.104.9.2763}.
%Type = Article
\bibitem[{Andor et~al.(2021)Andor, Fink, Frondel, Gerster and
  Horvath}]{andor_kostenloser_2021}
\bibinfo{author}{Andor, M.A.}, \bibinfo{author}{Fink, L.},
  \bibinfo{author}{Frondel, M.}, \bibinfo{author}{Gerster, A.},
  \bibinfo{author}{Horvath, M.}, \bibinfo{year}{2021}.
\newblock \bibinfo{title}{Kostenloser Ö{PNV}: {Akzeptanz} in der
  {Bevölkerung} und mögliche {Auswirkungen} auf das {Mobilitätsverhalten}}.
\newblock \bibinfo{journal}{List Forum für Wirtschafts- und Finanzpolitik}
  \bibinfo{volume}{46}, \bibinfo{pages}{299--325}.
\newblock \URLprefix
  \url{https://link.springer.com/10.1007/s41025-020-00207-y},
  \DOIprefix\doi{10.1007/s41025-020-00207-y}.
%Type = Article
\bibitem[{Axsen et~al.(2020)Axsen, Plötz and Wolinetz}]{axsen_crafting_2020}
\bibinfo{author}{Axsen, J.}, \bibinfo{author}{Plötz, P.},
  \bibinfo{author}{Wolinetz, M.}, \bibinfo{year}{2020}.
\newblock \bibinfo{title}{Crafting strong, integrated policy mixes for deep
  {CO2} mitigation in road transport}.
\newblock \bibinfo{journal}{Nature Climate Change} \bibinfo{volume}{10},
  \bibinfo{pages}{809--818}.
\newblock \URLprefix \url{http://dx.doi.org/10.1038/s41558-020-0877-y},
  \DOIprefix\doi{10.1038/s41558-020-0877-y}. \bibinfo{note}{publisher: Springer
  US}.
%Type = Article
\bibitem[{Banister(2008)}]{banister_sustainable_2008}
\bibinfo{author}{Banister, D.}, \bibinfo{year}{2008}.
\newblock \bibinfo{title}{The sustainable mobility paradigm}.
\newblock \bibinfo{journal}{Transport Policy} \bibinfo{volume}{15},
  \bibinfo{pages}{73--80}.
\newblock \URLprefix
  \url{https://linkinghub.elsevier.com/retrieve/pii/S0967070X07000820},
  \DOIprefix\doi{10.1016/j.tranpol.2007.10.005}.
%Type = Article
\bibitem[{Barnea et~al.(2022)Barnea, Hagemann and Wurster}]{barnea_policy_2022}
\bibinfo{author}{Barnea, G.}, \bibinfo{author}{Hagemann, C.},
  \bibinfo{author}{Wurster, S.}, \bibinfo{year}{2022}.
\newblock \bibinfo{title}{Policy instruments matter: {Support} schemes for
  renewable energy capacity in worldwide comparison}.
\newblock \bibinfo{journal}{Energy Policy} \bibinfo{volume}{168},
  \bibinfo{pages}{113093}.
\newblock \URLprefix
  \url{https://linkinghub.elsevier.com/retrieve/pii/S0301421522003184},
  \DOIprefix\doi{10.1016/j.enpol.2022.113093}.
%Type = Article
\bibitem[{Basso and Silva(2014)}]{basso_efficiency_2014}
\bibinfo{author}{Basso, L.J.}, \bibinfo{author}{Silva, H.E.},
  \bibinfo{year}{2014}.
\newblock \bibinfo{title}{Efficiency and {Substitutability} of {Transit}
  {Subsidies} and {Other} {Urban} {Transport} {Policies}}.
\newblock \bibinfo{journal}{American Economic Journal: Economic Policy}
  \bibinfo{volume}{6}, \bibinfo{pages}{1--33}.
\newblock \URLprefix \url{https://pubs.aeaweb.org/doi/10.1257/pol.6.4.1},
  \DOIprefix\doi{10.1257/pol.6.4.1}.
%Type = Article
\bibitem[{Bauernschuster et~al.(2017)Bauernschuster, Hener and
  Rainer}]{bauernschuster_when_2017}
\bibinfo{author}{Bauernschuster, S.}, \bibinfo{author}{Hener, T.},
  \bibinfo{author}{Rainer, H.}, \bibinfo{year}{2017}.
\newblock \bibinfo{title}{When {Labor} {Disputes} {Bring} {Cities} to a
  {Standstill}: {The} {Impact} of {Public} {Transit} {Strikes} on {Traffic},
  {Accidents}, {Air} {Pollution}, and {Health}}.
\newblock \bibinfo{journal}{American Economic Journal: Economic Policy}
  \bibinfo{volume}{9}, \bibinfo{pages}{1--37}.
\newblock \URLprefix \url{https://pubs.aeaweb.org/doi/10.1257/pol.20150414},
  \DOIprefix\doi{10.1257/pol.20150414}.
%Type = Article
\bibitem[{Becker et~al.(2017)Becker, Loder, Schmid and
  Axhausen}]{becker_modeling_2017}
\bibinfo{author}{Becker, H.}, \bibinfo{author}{Loder, A.},
  \bibinfo{author}{Schmid, B.}, \bibinfo{author}{Axhausen, K.W.},
  \bibinfo{year}{2017}.
\newblock \bibinfo{title}{Modeling car-sharing membership as a mobility tool:
  {A} multivariate {Probit} approach with latent variables}.
\newblock \bibinfo{journal}{Travel Behaviour and Society} \bibinfo{volume}{8},
  \bibinfo{pages}{26--36}.
\newblock \URLprefix \url{http://dx.doi.org/10.1016/j.tbs.2017.04.006},
  \DOIprefix\doi{10.1016/j.tbs.2017.04.006}.
%Type = Book
\bibitem[{Ben-Akiva and Lerman(1985)}]{ben-akiva_discrete_1985}
\bibinfo{author}{Ben-Akiva, M.E.}, \bibinfo{author}{Lerman, S.R.},
  \bibinfo{year}{1985}.
\newblock \bibinfo{title}{Discrete choice analysis: theory and application to
  travel demand}.
\newblock \bibinfo{publisher}{MIT press}, \bibinfo{address}{Cambridge, MA.}
%Type = Article
\bibitem[{Buehler et~al.(2016)Buehler, Pucher, Gerike and
  Götschi}]{buehler_reducing_2016}
\bibinfo{author}{Buehler, R.}, \bibinfo{author}{Pucher, J.},
  \bibinfo{author}{Gerike, R.}, \bibinfo{author}{Götschi, T.},
  \bibinfo{year}{2016}.
\newblock \bibinfo{title}{Reducing car dependence in the heart of {Europe}:
  lessons from {Germany}, {Austria}, and {Switzerland}}.
\newblock \bibinfo{journal}{Transport Reviews} \bibinfo{volume}{37},
  \bibinfo{pages}{4--28}.
\newblock \DOIprefix\doi{10.1080/01441647.2016.1177799}.
%Type = Article
\bibitem[{Bull et~al.(2021)Bull, Muñoz and Silva}]{bull_impact_2021}
\bibinfo{author}{Bull, O.}, \bibinfo{author}{Muñoz, J.C.},
  \bibinfo{author}{Silva, H.E.}, \bibinfo{year}{2021}.
\newblock \bibinfo{title}{The impact of fare-free public transport on travel
  behavior: {Evidence} from a randomized controlled trial}.
\newblock \bibinfo{journal}{Regional Science and Urban Economics}
  \bibinfo{volume}{86}, \bibinfo{pages}{103616}.
\newblock \URLprefix
  \url{https://linkinghub.elsevier.com/retrieve/pii/S016604622030301X},
  \DOIprefix\doi{10.1016/j.regsciurbeco.2020.103616}.
%Type = Techreport
\bibitem[{Bundesfinanzministerium(2022)}]{bundesfinanzministerium_masnahmenpaket_2022}
\bibinfo{author}{Bundesfinanzministerium}, \bibinfo{year}{2022}.
\newblock \bibinfo{title}{Maßnahmenpaket des {Bundes} zum {Umgang} mit den
  hohen {Energiekosten}}.
\newblock \bibinfo{type}{Ergebnis des {Koalitionsausschusses}}.
\newblock \URLprefix
  \url{https://www.bundesfinanzministerium.de/Content/DE/Downloads/2022-03-23-massnahmenpaket-bund-hohe-energiekosten.pdf?__blob=publicationFile&v=6}.
%Type = Techreport
\bibitem[{{Bundesministerium f{\"u}r Verkehr und digitale
  Infrastruktur}(2018a)}]{mid2017}
\bibinfo{author}{{Bundesministerium f{\"u}r Verkehr und digitale
  Infrastruktur}}, \bibinfo{year}{2018}a.
\newblock \bibinfo{title}{{Mobilit{\"a}t in Deutschland Ergebnisbericht}}.
\newblock \bibinfo{type}{Technical Report}. \bibinfo{address}{Bonn}.
\newblock \URLprefix
  \url{https://www.bmvi.de/SharedDocs/DE/Artikel/G/mobilitaet-in-deutschland.html}.
%Type = Techreport
\bibitem[{{Bundesministerium f{\"u}r Verkehr und digitale
  Infrastruktur}(2018b)}]{BundesministeriumfurVerkehrunddigitaleInfrastruktur.2018}
\bibinfo{author}{{Bundesministerium f{\"u}r Verkehr und digitale
  Infrastruktur}}, \bibinfo{year}{2018}b.
\newblock \bibinfo{title}{{Mobilit{\"a}t in Deutschland Ergebnisbericht}}.
\newblock \bibinfo{type}{Technical Report}. \bibinfo{address}{Bonn}.
\newblock \URLprefix
  \url{https://www.bmvi.de/SharedDocs/DE/Artikel/G/mobilitaet-in-deutschland.html}.
%Type = Article
\bibitem[{{Carbajo}(1988)}]{carbajo_economics_1988}
\bibinfo{author}{{Carbajo}}, \bibinfo{year}{1988}.
\newblock \bibinfo{title}{The {Economics} of {Travel} {Passes}: {Non}-{Uniform}
  {Pricing} in {Transport}}.
\newblock \bibinfo{journal}{Journal of Transport Economics and Policy}
  \bibinfo{volume}{22}, \bibinfo{pages}{153--173}.
\newblock \DOIprefix\doi{10.2307/20052843}.
%Type = Article
\bibitem[{Carr and Hesse(2020)}]{carr_mobility_2020}
\bibinfo{author}{Carr, C.}, \bibinfo{author}{Hesse, M.}, \bibinfo{year}{2020}.
\newblock \bibinfo{title}{Mobility policy through the lens of policy mobility:
  {The} post-political case of introducing free transit in {Luxembourg}}.
\newblock \bibinfo{journal}{Journal of Transport Geography}
  \bibinfo{volume}{83}, \bibinfo{pages}{102634}.
\newblock \URLprefix
  \url{https://linkinghub.elsevier.com/retrieve/pii/S0966692319304892},
  \DOIprefix\doi{10.1016/j.jtrangeo.2020.102634}.
%Type = Article
\bibitem[{Cats et~al.(2014)Cats, Reimal and Susilo}]{cats_public_2014}
\bibinfo{author}{Cats, O.}, \bibinfo{author}{Reimal, T.},
  \bibinfo{author}{Susilo, Y.}, \bibinfo{year}{2014}.
\newblock \bibinfo{title}{Public {Transport} {Pricing} {Policy} - {Empirical}
  {Evidence} from a {Fare}-{Free} {Scheme} in {Tallinn}, {Estonia}}.
\newblock \bibinfo{journal}{Transportation Research Record}
  \bibinfo{volume}{2415}, \bibinfo{pages}{89--96}.
\newblock \URLprefix \url{http://trrjournalonline.trb.org/doi/10.3141/2415-10},
  \DOIprefix\doi{10.3141/2415-10}.
%Type = Article
\bibitem[{Christidis et~al.(2022)Christidis, Navajas~Cawood and
  Fiorello}]{christidis_challenges_2022}
\bibinfo{author}{Christidis, P.}, \bibinfo{author}{Navajas~Cawood, E.},
  \bibinfo{author}{Fiorello, D.}, \bibinfo{year}{2022}.
\newblock \bibinfo{title}{Challenges for urban transport policy after the
  {Covid}-19 pandemic: {Main} findings from a survey in 20 {European} cities}.
\newblock \bibinfo{journal}{Transport Policy} \bibinfo{volume}{129},
  \bibinfo{pages}{105--116}.
\newblock \URLprefix
  \url{https://linkinghub.elsevier.com/retrieve/pii/S0967070X2200292X},
  \DOIprefix\doi{10.1016/j.tranpol.2022.10.007}.
%Type = Article
\bibitem[{Djeffal et~al.(2022)Djeffal, Siewert and Wurster}]{djeffal_role_2022}
\bibinfo{author}{Djeffal, C.}, \bibinfo{author}{Siewert, M.B.},
  \bibinfo{author}{Wurster, S.}, \bibinfo{year}{2022}.
\newblock \bibinfo{title}{Role of the state and responsibility in governing
  artificial intelligence: a comparative analysis of {AI} strategies}.
\newblock \bibinfo{journal}{Journal of European Public Policy}
  \bibinfo{volume}{29}, \bibinfo{pages}{1799--1821}.
\newblock \URLprefix
  \url{https://www.tandfonline.com/doi/full/10.1080/13501763.2022.2094987},
  \DOIprefix\doi{10.1080/13501763.2022.2094987}.
%Type = Article
\bibitem[{Eisenmann et~al.(2021)Eisenmann, Nobis, Kolarova, Lenz and
  Winkler}]{eisenmann_transport_2021}
\bibinfo{author}{Eisenmann, C.}, \bibinfo{author}{Nobis, C.},
  \bibinfo{author}{Kolarova, V.}, \bibinfo{author}{Lenz, B.},
  \bibinfo{author}{Winkler, C.}, \bibinfo{year}{2021}.
\newblock \bibinfo{title}{Transport mode use during the {COVID}-19 lockdown
  period in {Germany}: {The} car became more important, public transport lost
  ground}.
\newblock \bibinfo{journal}{Transport Policy} \bibinfo{volume}{103},
  \bibinfo{pages}{60--67}.
\newblock \URLprefix
  \url{https://linkinghub.elsevier.com/retrieve/pii/S0967070X21000184},
  \DOIprefix\doi{10.1016/j.tranpol.2021.01.012}.
%Type = Misc
\bibitem[{{EuroStat}(2023)}]{eurostat2023}
\bibinfo{author}{{EuroStat}}, \bibinfo{year}{2023}.
\newblock \bibinfo{title}{Population on 1 january by age, sex and nuts 2
  region}.
\newblock
  \bibinfo{note}{\url{https://ec.europa.eu/eurostat/databrowser/product/page/demo_r_d2jan}}.
%Type = Article
\bibitem[{Falk et~al.(2018)Falk, Becker, Dohmen, Enke, Huffman and
  Sunde}]{falk2018global}
\bibinfo{author}{Falk, A.}, \bibinfo{author}{Becker, A.},
  \bibinfo{author}{Dohmen, T.}, \bibinfo{author}{Enke, B.},
  \bibinfo{author}{Huffman, D.}, \bibinfo{author}{Sunde, U.},
  \bibinfo{year}{2018}.
\newblock \bibinfo{title}{Global evidence on economic preferences}.
\newblock \bibinfo{journal}{The Quarterly Journal of Economics}
  \bibinfo{volume}{133}, \bibinfo{pages}{1645--1692}.
%Type = Article
\bibitem[{Falk et~al.(2022)Falk, Becker, Dohmen, Huffman and
  Sunde}]{falk2022preference}
\bibinfo{author}{Falk, A.}, \bibinfo{author}{Becker, A.},
  \bibinfo{author}{Dohmen, T.}, \bibinfo{author}{Huffman, D.},
  \bibinfo{author}{Sunde, U.}, \bibinfo{year}{2022}.
\newblock \bibinfo{title}{The preference survey module: A validated instrument
  for measuring risk, time, and social preferences}.
\newblock \bibinfo{journal}{Management Science} .
%Type = Article
\bibitem[{Gaus et~al.(2023)Gaus, Murray and Link}]{gaus_9-euro-ticket_2023}
\bibinfo{author}{Gaus, D.}, \bibinfo{author}{Murray, N.},
  \bibinfo{author}{Link, H.}, \bibinfo{year}{2023}.
\newblock \bibinfo{title}{9-{Euro}-{Ticket}: {Niedrigere} {Preise} allein
  stärken {Alltagsmobilität} mit öffentlichen {Verkehrsmitteln} nicht}.
\newblock \bibinfo{journal}{DIW Wochenbericht} \URLprefix
  \url{http://www.diw.de/sixcms/detail.php?id=diw_01.c.869729.de},
  \DOIprefix\doi{10.18723/DIW_WB:2023-14-1}. \bibinfo{note}{publisher: DIW -
  Deutsches Institut für Wirtschaftsforschung Version Number: 2.0}.
%Type = Article
\bibitem[{Gerike and Schulz(2018)}]{Gerike.2018}
\bibinfo{author}{Gerike, R.}, \bibinfo{author}{Schulz, A.},
  \bibinfo{year}{2018}.
\newblock \bibinfo{title}{Workshop synthesis: Surveys on long-distance travel
  and other rare events}.
\newblock \bibinfo{journal}{Transportation Research Procedia}
  \bibinfo{volume}{32}, \bibinfo{pages}{535--541}.
\newblock \DOIprefix\doi{10.1016/j. trpro.2018.10.032}.
%Type = Article
\bibitem[{González et~al.(2008)González, Hidalgo and
  Barabási}]{gonzalez_understanding_2008}
\bibinfo{author}{González, M.C.}, \bibinfo{author}{Hidalgo, C.A.},
  \bibinfo{author}{Barabási, A.L.}, \bibinfo{year}{2008}.
\newblock \bibinfo{title}{Understanding individual human mobility patterns}.
\newblock \bibinfo{journal}{Nature} \bibinfo{volume}{453},
  \bibinfo{pages}{779--782}.
\newblock \DOIprefix\doi{10.1038/nature06958}.
%Type = Article
\bibitem[{Greening et~al.(2000)Greening, Greene and
  Difiglio}]{greening_energy_2000}
\bibinfo{author}{Greening, L.A.}, \bibinfo{author}{Greene, D.L.},
  \bibinfo{author}{Difiglio, C.}, \bibinfo{year}{2000}.
\newblock \bibinfo{title}{Energy efficiency and consumption — the rebound
  effect — a survey}.
\newblock \bibinfo{journal}{Energy Policy} \bibinfo{volume}{28},
  \bibinfo{pages}{389--401}.
\newblock \DOIprefix\doi{10.1016/S0301-4215(00)00021-5}.
%Type = Article
\bibitem[{Heinen et~al.(2010)Heinen, van Wee and
  Maat}]{Heinen_Bike_Temperature}
\bibinfo{author}{Heinen, E.}, \bibinfo{author}{van Wee, B.},
  \bibinfo{author}{Maat, K.}, \bibinfo{year}{2010}.
\newblock \bibinfo{title}{Commuting by bicycle: An overview of the literature}.
\newblock \bibinfo{journal}{Transport Reviews} \bibinfo{volume}{30},
  \bibinfo{pages}{59--96}.
\newblock \URLprefix \url{https://doi.org/10.1080/01441640903187001},
  \DOIprefix\doi{10.1080/01441640903187001},
  \href{http://arxiv.org/abs/https://doi.org/10.1080/01441640903187001}{{\tt
  arXiv:https://doi.org/10.1080/01441640903187001}}.
%Type = Article
\bibitem[{Hoppmann et~al.(2014)Hoppmann, Huenteler and
  Girod}]{hoppmann_compulsive_2014}
\bibinfo{author}{Hoppmann, J.}, \bibinfo{author}{Huenteler, J.},
  \bibinfo{author}{Girod, B.}, \bibinfo{year}{2014}.
\newblock \bibinfo{title}{Compulsive policy-making—{The} evolution of the
  {German} feed-in tariff system for solar photovoltaic power}.
\newblock \bibinfo{journal}{Research Policy} \bibinfo{volume}{43},
  \bibinfo{pages}{1422--1441}.
\newblock \URLprefix
  \url{https://linkinghub.elsevier.com/retrieve/pii/S0048733314000249},
  \DOIprefix\doi{10.1016/j.respol.2014.01.014}.
%Type = Article
\bibitem[{Hymel et~al.(2010)Hymel, Small and Dender}]{hymel_induced_2010}
\bibinfo{author}{Hymel, K.M.}, \bibinfo{author}{Small, K.A.},
  \bibinfo{author}{Dender, K.V.}, \bibinfo{year}{2010}.
\newblock \bibinfo{title}{Induced demand and rebound effects in road
  transport}.
\newblock \bibinfo{journal}{Transportation Research Part B: Methodological}
  \bibinfo{volume}{44}, \bibinfo{pages}{1220--1241}.
\newblock \DOIprefix\doi{10.1016/J.TRB.2010.02.007}.
%Type = Article
\bibitem[{Keblowski(2020)}]{keblowski_why_2020}
\bibinfo{author}{Keblowski, W.}, \bibinfo{year}{2020}.
\newblock \bibinfo{title}{Why (not) abolish fares? {Exploring} the global
  geography of fare-free public transport}.
\newblock \bibinfo{journal}{Transportation} \bibinfo{volume}{47},
  \bibinfo{pages}{2807--2835}.
\newblock \URLprefix \url{http://link.springer.com/10.1007/s11116-019-09986-6},
  \DOIprefix\doi{10.1007/s11116-019-09986-6}.
%Type = Article
\bibitem[{Knoflacher(2019)}]{knoflacher_wem_2019}
\bibinfo{author}{Knoflacher, H.}, \bibinfo{year}{2019}.
\newblock \bibinfo{title}{Wem gehört die {Stadt}? {Plädoyer} für einen
  {Paradigmenwechsel} in {Stadtplanung} und {Verkehrspolitik}}.
\newblock \bibinfo{journal}{DER NAHVERKEHR} \bibinfo{volume}{37},
  \bibinfo{pages}{51}.
%Type = Article
\bibitem[{Kolarova et~al.(2021)Kolarova, Eisenmann, Nobis, Winkler and
  Lenz}]{kolarova_analysing_2021}
\bibinfo{author}{Kolarova, V.}, \bibinfo{author}{Eisenmann, C.},
  \bibinfo{author}{Nobis, C.}, \bibinfo{author}{Winkler, C.},
  \bibinfo{author}{Lenz, B.}, \bibinfo{year}{2021}.
\newblock \bibinfo{title}{Analysing the impact of the {COVID}-19 outbreak on
  everyday travel behaviour in {Germany} and potential implications for future
  travel patterns}.
\newblock \bibinfo{journal}{European Transport Research Review}
  \bibinfo{volume}{13}, \bibinfo{pages}{27}.
\newblock \URLprefix
  \url{https://etrr.springeropen.com/articles/10.1186/s12544-021-00486-2},
  \DOIprefix\doi{10.1186/s12544-021-00486-2}.
%Type = Article
\bibitem[{Krämer et~al.(2022)Krämer, Wilger and
  Bongaerts}]{kramer_9-euro-ticket_2022}
\bibinfo{author}{Krämer, A.}, \bibinfo{author}{Wilger, G.},
  \bibinfo{author}{Bongaerts, R.}, \bibinfo{year}{2022}.
\newblock \bibinfo{title}{Das 9-{Euro}-{Ticket}: {Erfahrungen},
  {Wirkungsmechanismen} und {Nachfolgeangebot}}.
\newblock \bibinfo{journal}{Wirtschaftsdienst} \bibinfo{volume}{102},
  \bibinfo{pages}{873--879}.
\newblock \URLprefix \url{https://link.springer.com/10.1007/s10273-022-3313-2},
  \DOIprefix\doi{10.1007/s10273-022-3313-2}.
%Type = Article
\bibitem[{Köllinger(2022)}]{kollinger_cascais_2022}
\bibinfo{author}{Köllinger}, \bibinfo{year}{2022}.
\newblock \bibinfo{title}{Cascais free public transit services result in 10\%
  more users} \URLprefix
  \url{https://www.eltis.org/in-brief/news/cascais-free-public-transit-services-result-10-more-users}.
  \bibinfo{note}{https://www.themayor.eu/en/a/view/two-years-of-free-public-transport-in-cascais-have-drawn-in-10-more-commuters-10052}.
%Type = Article
\bibitem[{Larcom et~al.(2017)Larcom, Rauch and Willems}]{larcom_benefits_2017}
\bibinfo{author}{Larcom, S.}, \bibinfo{author}{Rauch, F.},
  \bibinfo{author}{Willems, T.}, \bibinfo{year}{2017}.
\newblock \bibinfo{title}{The {Benefits} of {Forced} {Experimentation}:
  {Striking} {Evidence} from the {London} {Underground} {Network}*}.
\newblock \bibinfo{journal}{The Quarterly Journal of Economics}
  \bibinfo{volume}{132}, \bibinfo{pages}{2019--2055}.
\newblock \URLprefix
  \url{https://academic.oup.com/qje/article/132/4/2019/3857744},
  \DOIprefix\doi{10.1093/qje/qjx020}.
%Type = Article
\bibitem[{Li and Hensher(2011)}]{li_crowding_2011}
\bibinfo{author}{Li, Z.}, \bibinfo{author}{Hensher, D.A.},
  \bibinfo{year}{2011}.
\newblock \bibinfo{title}{Crowding and public transport: {A} review of
  willingness to pay evidence and its relevance in project appraisal}.
\newblock \bibinfo{journal}{Transport Policy} \bibinfo{volume}{18},
  \bibinfo{pages}{880--887}.
\newblock \DOIprefix\doi{10.1016/J.TRANPOL.2011.06.003}.
  \bibinfo{note}{publisher: Pergamon}.
%Type = Article
\bibitem[{Lindsey and Santos(2020)}]{lindsey_addressing_2020}
\bibinfo{author}{Lindsey, R.}, \bibinfo{author}{Santos, G.},
  \bibinfo{year}{2020}.
\newblock \bibinfo{title}{Addressing transportation and environmental
  externalities with economics: {Are} policy makers listening?}
\newblock \bibinfo{journal}{Research in Transportation Economics}
  \bibinfo{volume}{82}, \bibinfo{pages}{100872}.
\newblock \DOIprefix\doi{10.1016/j.retrec.2020.100872}.
%Type = Article
\bibitem[{Loder et~al.(2022)Loder, Bliemer and Axhausen}]{loder_optimal_2022}
\bibinfo{author}{Loder, A.}, \bibinfo{author}{Bliemer, M.C.},
  \bibinfo{author}{Axhausen, K.W.}, \bibinfo{year}{2022}.
\newblock \bibinfo{title}{Optimal pricing and investment in a multi-modal city
  — {Introducing} a macroscopic network design problem based on the {MFD}}.
\newblock \bibinfo{journal}{Transportation Research Part A: Policy and
  Practice} \bibinfo{volume}{156}, \bibinfo{pages}{113--132}.
\newblock \URLprefix \url{https://doi.org/10.1016/j.tra.2021.11.026},
  \DOIprefix\doi{10.1016/j.tra.2021.11.026}.
%Type = Article
\bibitem[{Lugtig et~al.(2022)Lugtig, Roth and
  Schouten}]{lugtig_nonresponse_2022}
\bibinfo{author}{Lugtig, P.}, \bibinfo{author}{Roth, K.},
  \bibinfo{author}{Schouten, B.}, \bibinfo{year}{2022}.
\newblock \bibinfo{title}{Nonresponse analysis in a longitudinal
  smartphone-based travel study}.
\newblock \bibinfo{journal}{Survey Research Methods} , \bibinfo{pages}{13--27
  Pages}\URLprefix \url{https://ojs.ub.uni-konstanz.de/srm/article/view/7835},
  \DOIprefix\doi{10.18148/SRM/2022.V16I1.7835}. \bibinfo{note}{artwork Size:
  13-27 Pages Publisher: Survey Research Methods}.
%Type = Misc
\bibitem[{civity
  Management~Consultants(2021)}]{civity_management_consultants_gutachten_2021}
\bibinfo{author}{civity Management~Consultants}, \bibinfo{year}{2021}.
\newblock \bibinfo{title}{Gutachten 365-{Euro}-{Ticket} für {Alle} -
  {Endbericht}}.
\newblock \URLprefix
  \url{https://www.vgn.de/5ef141c9-e1c4-655e-a3de-196f385b0e70}.
%Type = Article
\bibitem[{Molloy et~al.(2022)Molloy, Castro, Götschi, Schoeman, Tchervenkov,
  Tomic, Hintermann and Axhausen}]{molloy_mobis_2022}
\bibinfo{author}{Molloy, J.}, \bibinfo{author}{Castro, A.},
  \bibinfo{author}{Götschi, T.}, \bibinfo{author}{Schoeman, B.},
  \bibinfo{author}{Tchervenkov, C.}, \bibinfo{author}{Tomic, U.},
  \bibinfo{author}{Hintermann, B.}, \bibinfo{author}{Axhausen, K.W.},
  \bibinfo{year}{2022}.
\newblock \bibinfo{title}{The {MOBIS} dataset: a large {GPS} dataset of
  mobility behaviour in {Switzerland}}.
\newblock \bibinfo{journal}{Transportation} \URLprefix
  \url{https://link.springer.com/10.1007/s11116-022-10299-4},
  \DOIprefix\doi{10.1007/s11116-022-10299-4}.
%Type = Article
\bibitem[{Molloy et~al.(2021)Molloy, Schatzmann, Schoeman, Tchervenkov,
  Hintermann and Axhausen}]{molloy_observed_2021}
\bibinfo{author}{Molloy, J.}, \bibinfo{author}{Schatzmann, T.},
  \bibinfo{author}{Schoeman, B.}, \bibinfo{author}{Tchervenkov, C.},
  \bibinfo{author}{Hintermann, B.}, \bibinfo{author}{Axhausen, K.W.},
  \bibinfo{year}{2021}.
\newblock \bibinfo{title}{Observed impacts of the {Covid}-19 first wave on
  travel behaviour in {Switzerland} based on a large {GPS} panel}.
\newblock \bibinfo{journal}{Transport Policy} \bibinfo{volume}{104},
  \bibinfo{pages}{43--51}.
\newblock \DOIprefix\doi{10.1016/J.TRANPOL.2021.01.009}.
%Type = Inproceedings
\bibitem[{Paget-Seekins(2023)}]{paget2023}
\bibinfo{author}{Paget-Seekins, L.}, \bibinfo{year}{2023}.
\newblock \bibinfo{title}{{Transit Ridership and Fare Policy in the Recovery
  from COVID: A Scan of Industry Responses and Plans}},
  \bibinfo{organization}{Paper presented at the 102nd Annual Meeting of the
  Transportation Research Board}, \bibinfo{address}{Washington, DC}.
%Type = Article
\bibitem[{Parry et~al.(2007)Parry, Walls and
  Harrington}]{parry_automobile_2007}
\bibinfo{author}{Parry, I.W.H.}, \bibinfo{author}{Walls, M.},
  \bibinfo{author}{Harrington, W.}, \bibinfo{year}{2007}.
\newblock \bibinfo{title}{Automobile {Externalities} and {Policies}}.
\newblock \bibinfo{journal}{Journal of Economic Literature}
  \bibinfo{volume}{45}, \bibinfo{pages}{373--399}.
\newblock \DOIprefix\doi{10.2139/ssrn.927794}.
%Type = Article
\bibitem[{Prud'homme et~al.(2012)Prud'homme, Koning, Lenormand and
  Fehr}]{prudhomme_public_2012}
\bibinfo{author}{Prud'homme, R.}, \bibinfo{author}{Koning, M.},
  \bibinfo{author}{Lenormand, L.}, \bibinfo{author}{Fehr, A.},
  \bibinfo{year}{2012}.
\newblock \bibinfo{title}{Public transport congestion costs: {The} case of the
  {Paris} subway}.
\newblock \bibinfo{journal}{Transport Policy} \bibinfo{volume}{21},
  \bibinfo{pages}{101--109}.
\newblock \URLprefix
  \url{https://linkinghub.elsevier.com/retrieve/pii/S0967070X11001302},
  \DOIprefix\doi{10.1016/j.tranpol.2011.11.002}.
%Type = Article
\bibitem[{Santos et~al.(2010)Santos, Behrendt, Maconi, Shirvani and
  Teytelboym}]{Santos2010}
\bibinfo{author}{Santos, G.}, \bibinfo{author}{Behrendt, H.},
  \bibinfo{author}{Maconi, L.}, \bibinfo{author}{Shirvani, T.},
  \bibinfo{author}{Teytelboym, A.}, \bibinfo{year}{2010}.
\newblock \bibinfo{title}{Part {I}: {Externalities} and economic policies in
  road transport}.
\newblock \bibinfo{journal}{Research in Transportation Economics}
  \bibinfo{volume}{28}, \bibinfo{pages}{2--45}.
\newblock \DOIprefix\doi{10.1016/j.retrec.2009.11.002}.
%Type = Article
\bibitem[{Sharaby and Shiftan(2012)}]{sharaby_impact_2012}
\bibinfo{author}{Sharaby, N.}, \bibinfo{author}{Shiftan, Y.},
  \bibinfo{year}{2012}.
\newblock \bibinfo{title}{The impact of fare integration on travel behavior and
  transit ridership}.
\newblock \bibinfo{journal}{Transport Policy} \bibinfo{volume}{21},
  \bibinfo{pages}{63--70}.
\newblock \URLprefix
  \url{https://linkinghub.elsevier.com/retrieve/pii/S0967070X12000169},
  \DOIprefix\doi{10.1016/j.tranpol.2012.01.015}.
%Type = Article
\bibitem[{Shoup(1997)}]{shoup_high_1997}
\bibinfo{author}{Shoup, D.C.}, \bibinfo{year}{1997}.
\newblock \bibinfo{title}{The {High} {Cost} of {Free} {Parking}}.
\newblock \bibinfo{journal}{Journal of Planning Education and Research}
  \bibinfo{volume}{17}, \bibinfo{pages}{3--20}.
\newblock \DOIprefix\doi{10.1177/016344300022005001}.
%Type = Book
\bibitem[{Small and Verhoef(2007)}]{small_economics_2007}
\bibinfo{author}{Small, K.A.}, \bibinfo{author}{Verhoef, E.T.},
  \bibinfo{year}{2007}.
\newblock \bibinfo{title}{The economics of urban transportation}.
\newblock \bibinfo{publisher}{Routledge}, \bibinfo{address}{London}.
\newblock \bibinfo{note}{ISSN: 978-0-415-28514-8}.
%Type = Article
\bibitem[{Spence(1977)}]{spence_nonlinear_1977}
\bibinfo{author}{Spence, M.}, \bibinfo{year}{1977}.
\newblock \bibinfo{title}{Nonlinear prices and welfare}.
\newblock \bibinfo{journal}{Journal of Public Economics} \bibinfo{volume}{8},
  \bibinfo{pages}{1--18}.
\newblock \URLprefix
  \url{https://linkinghub.elsevier.com/retrieve/pii/0047272777900251},
  \DOIprefix\doi{10.1016/0047-2727(77)90025-1}.
%Type = Article
\bibitem[{Storchmann(2003)}]{storchmann_externalities_2003}
\bibinfo{author}{Storchmann, K.}, \bibinfo{year}{2003}.
\newblock \bibinfo{title}{Externalities by {Automobiles} and {Fare}-{Free}
  {Transit} in {Germany} — {A} {Paradigm} {Shift}?}
\newblock \bibinfo{journal}{Journal of Public Transportation}
  \bibinfo{volume}{6}, \bibinfo{pages}{89--105}.
\newblock \URLprefix \url{http://scholarcommons.usf.edu/jpt/vol6/iss4/5/},
  \DOIprefix\doi{10.5038/2375-0901.6.4.5}.
%Type = Article
\bibitem[{Tirachini et~al.(2014)Tirachini, Hensher and
  Rose}]{tirachini_multimodal_2014}
\bibinfo{author}{Tirachini, A.}, \bibinfo{author}{Hensher, D.A.},
  \bibinfo{author}{Rose, J.M.}, \bibinfo{year}{2014}.
\newblock \bibinfo{title}{Multimodal pricing and optimal design of urban public
  transport: {The} interplay between traffic congestion and bus crowding}.
\newblock \bibinfo{journal}{Transportation Research Part B: Methodological}
  \bibinfo{volume}{61}, \bibinfo{pages}{33--54}.
\newblock \DOIprefix\doi{10.1016/j.trb.2014.01.003}.
%Type = Article
\bibitem[{Topp(1993)}]{topp_parking_1993}
\bibinfo{author}{Topp, H.H.}, \bibinfo{year}{1993}.
\newblock \bibinfo{title}{Parking policies to reduce car traffic in {German}
  cities}.
\newblock \bibinfo{journal}{Transport Reviews} \bibinfo{volume}{13},
  \bibinfo{pages}{83--95}.
\newblock \URLprefix
  \url{http://www.tandfonline.com/doi/abs/10.1080/01441649308716836},
  \DOIprefix\doi{10.1080/01441649308716836}.
%Type = Article
\bibitem[{Wardman and Murphy(2015)}]{wardman_passengers_2015}
\bibinfo{author}{Wardman, M.}, \bibinfo{author}{Murphy, P.},
  \bibinfo{year}{2015}.
\newblock \bibinfo{title}{Passengers’ valuations of train seating layout,
  position and occupancy}.
\newblock \bibinfo{journal}{Transportation Research Part A: Policy and
  Practice} \bibinfo{volume}{74}, \bibinfo{pages}{222--238}.
\newblock \URLprefix
  \url{https://linkinghub.elsevier.com/retrieve/pii/S0965856415000154},
  \DOIprefix\doi{10.1016/j.tra.2015.01.007}.
%Type = Article
\bibitem[{Weis and Axhausen(2009)}]{weis_induced_2009}
\bibinfo{author}{Weis, C.}, \bibinfo{author}{Axhausen, K.W.},
  \bibinfo{year}{2009}.
\newblock \bibinfo{title}{Induced travel demand: {Evidence} from a pseudo panel
  data based structural equations model}.
\newblock \bibinfo{journal}{Research in Transportation Economics}
  \bibinfo{volume}{25}, \bibinfo{pages}{8--18}.
\newblock \URLprefix
  \url{http://www.sciencedirect.com/science/article/pii/S0739885909000328},
  \DOIprefix\doi{https://doi.org/10.1016/j.retrec.2009.08.007}.
%Type = Article
\bibitem[{White(1984)}]{white_development_1984}
\bibinfo{author}{White, P.R.}, \bibinfo{year}{1984}.
\newblock \bibinfo{title}{Development of the ‘{Travelcard}’ {Concept} in
  {Urban} {Public} {Transport}}.
\newblock \bibinfo{journal}{The Service Industries Journal}
  \bibinfo{volume}{4}, \bibinfo{pages}{133--150}.
\newblock \DOIprefix\doi{10.1080/02642068400000067}.
%Type = Inproceedings
\bibitem[{{Wirtz, Matthias} et~al.(2015){Wirtz, Matthias}, {Vortisch, Peter}
  and {Chlond, Bastian}}]{wirtz_matthias_flatrate_2015}
\bibinfo{author}{{Wirtz, Matthias}}, \bibinfo{author}{{Vortisch, Peter}},
  \bibinfo{author}{{Chlond, Bastian}}, \bibinfo{year}{2015}.
\newblock \bibinfo{title}{Flatrate {Bias} in {Public} {Transportation}:
  {Magnitude} and {Reasoning}}, in: \bibinfo{booktitle}{Paper presented at the
  94th {Annual} {Meeting} of the {Transportation} {Research} {Board}},
  \bibinfo{address}{Washington D.C.}
%Type = Article
\bibitem[{Zhu et~al.(2010)Zhu, Levinson, Liu and Harder}]{zhu_traffic_2010}
\bibinfo{author}{Zhu, S.}, \bibinfo{author}{Levinson, D.},
  \bibinfo{author}{Liu, H.X.}, \bibinfo{author}{Harder, K.},
  \bibinfo{year}{2010}.
\newblock \bibinfo{title}{The traffic and behavioral effects of the {I}-{35W}
  {Mississippi} {River} bridge collapse}.
\newblock \bibinfo{journal}{Transportation Research Part A: Policy and
  Practice} \bibinfo{volume}{44}, \bibinfo{pages}{771--784}.
\newblock \DOIprefix\doi{10.1016/J.TRA.2010.07.001}.

\end{thebibliography}
